\documentclass[11pt,a4paper]{article}
\usepackage{jheppub_kim}
\usepackage{pdflscape}
\usepackage{amsmath}
\usepackage{amssymb}
\usepackage{dcolumn}
\usepackage{bm}
\usepackage{color}
\usepackage{epsfig}
\usepackage{amsfonts}
\usepackage{graphicx}
\usepackage{subfigure}
\usepackage{dcolumn}

\begin{document}

\title{Time Dependent Geometry  in  Massive Gravity}
\author[a]{Yaghoub Heydarzade,}
\author[b]{Prabir Rudra,}
\author[c]{Behnam Pourhassan,}
\author[d]{Mir Faizal,}
\author[e]{Ahmed Farag Ali,}
\author[f]{Farhad Darabi}

\affiliation[a] {Department of Physics, Azarbaijan Shahid Madani University, Tabriz, 53714-161, Iran}
\affiliation[b] {Department of Mathematics, Asutosh College, Kolkata-700 026, India}
\affiliation[c] {School of Physics, Damghan University, Damghan, 3671641167, Iran}
\affiliation[d] {Irving K. Barber School of Arts and Sciences,  University of British Columbia - Okanagan, Kelowna, BC V1V 1V7, Canada}
\affiliation[d] {Department of Physics and Astronomy, University of Lethbridge, Lethbridge,   Alberta, T1K 3M4, Canada  }
\affiliation[e] {Netherlands Institute for Advanced Study, Korte Spinhuissteeg 3, 1012 CG Amsterdam, Netherlands}
\affiliation[e] {Department of Physics, Faculty of Science, Benha University, Benha, 13518, Egypt}
\affiliation[f] {Department of Physics, Azarbaijan Shahid Madani University, Tabriz, 53714-161, Iran}

\emailAdd{heydarzade@azaruniv.edu}
\emailAdd{prudra.math@gmail.com}
\emailAdd{b.pourhassan@du.ac.ir}
\emailAdd{mirfaizalmir@googlemail.com}
\emailAdd{ahmed.ali@fsc.bu.edu.eg}
\emailAdd{f.darabi@azaruniv.edu (Corresponding author)}

\abstract{In this paper, we will analyze a time dependent geometry in a massive theory of gravity.
This will be done by analyzing
 Vaidya spacetime in such a massive theory of gravity.
As gravitational collapse is a time dependent system, we will
analyze it using the Vaidya spacetime in massive gravity. The
Vainshtein  and dRGT mechanisms are used to obtain a ghost free
massive gravity, and construct such time dependent solutions.
Singularities formed, their nature and strength will be studied in
detail. We will also study the thermodynamical aspects of such a
geometry by calculating the important thermodynamical quantities
for such a system, and analyzing the thermodynamical behavior of
such quantities.}


\maketitle
\section{Introduction}
The  observations from   type I supernovae indicate that our
universe is in a state of accelerated cosmic expansion \cite{super}-\cite{super5}.
This accelerated cosmic expansion can be explained by a cosmological constant term
in the Einstein equation, and the existence of such a cosmological constant is predicted
from  all quantum field theories. However, the cosmological constant paradigm
suffers from two well known problems as the ``cosmological constant problem'' and
the ``coincidence problem''.  These problems have motivated the people to
do research in the dark energy models \cite{DE}-\cite{DE2} and the modified theories
of gravity \cite{M1}. The latter theories are constrained by the solar system tests \cite{2a}-\cite{a2},
where the modifications have to occur only at the infrared limit.
It is possible to obtain an infrared modification of the general relativity
by  making the gravitons massive \cite{M2}, such that the small graviton mass
does not violate the known experimental bounds.
Even though this  has been done by adding a small Fierz-Pauli mass term to the original action of
general relativity  \cite{1}-\cite{2},  there are problems with the zero
mass limit of this theory due to the   force
mediated by the scalar graviton. Furthermore, such a modified theory of gravity
violates the experimental  bounds obtained from solar system experiments, and so it cannot be a physical theory
\cite{2a}-\cite{a2}.

It  was possible to resolve these problems
by using the   Vainshtein mechanism, which was based on the  inclusion of
non-linear
terms in the field equation \cite{4}-\cite{5}. Even though the Vainshtein mechanism
produces the general relativity in suitable limits, it contains higher derivative terms.
These higher derivative terms give rise to negative norm   Boulware-Deser ghosts \cite{6}.
This problem can also be resolved for a subclass of massive potentials, as it has been observed
that for such a subclass of massive potential the
Boulware-Deser ghosts do not appear \cite{7}-\cite{13}.
This has been done using dRGT mechanism, which is
a theory with one dynamical and one fixed metric
\cite{14}.  It is interesting to note that a mass term in the gravitational action can
also be generated from the spontaneous breaking of Lorentz symmetry at the cosmological scale.
\cite{lore}-\cite{lore2}.  Thus, the  massive gravity  might be produced by some interesting
theoretical considerations.

As massive gravity produces interesting deformation of the general relativity, it has
been used to study the behavior of various interesting systems.
The  black holes in Gauss-Bonnet massive gravity have been studied \cite{gaus}-\cite{gauss}, and it has been demonstrated
that the inclusion of mass term produces interesting deformation of these black holes.
The thermodynamics of such black holes has been studied in the extended phase space
\cite{M2}-\cite{phase}. It has been demonstrated
that the phase transition of black holes depends on the different parameters used in this
 massive gravity \cite{phase}.
Cosmological solutions with a well defined   initial values have been constructed in massive gravity \cite{cosm12}.
 The initial value constraints in massive gravity  have been used to study the
 spherically symmetric deformations of flat space,
 and it has been demonstrated that there is a physical sector of the theory, where the theory is stable \cite{cosm14}.

The massive theory of gravity has also been used to analyze the deformation of AdS spacetime,
and its CFT dual using the AdS/CFT correspondence \cite{ads}-\cite{cft}.
The holographic entanglement entropy of a field theory dual to the massive gravity has also
been studied \cite{ee}. It was observed using this holographic entanglement
entropy that  both  first order phase
transition  and  second order phase transition occur in this system.
The holographic complexity has also been calculated in the massive gravity \cite{cc}.
The  stability of solution in massive gravity have been studied using
holographic conductivity \cite{conduct}. Thus, massive gravity has  been used
to study interesting physical systems using gravity/gravity duality. This is another
motivation for analyzing solutions in massive gravity. As the massive gravity is interesting
modification of general relativity, we will analyze a time dependent solutions in massive gravity.

The  time dependent deformation of AdS solution has been used to
analyze the time dependent field theories
\cite{time}-\cite{time1}, and it has led some interests to study
such solutions in massive gravity. These solutions are obtained as
deformations of the Vaidya spacetime, which is a time dependent
spherically symmetric spacetime \cite{sphe}-\cite{sphe1}. In fact,
a time dependent black hole solution \cite{td12}, and  a time
dependent solutions in AdS/CFT correspondence \cite{td21}, have
been studied using massive gravity. The Vaidya spacetime has
already been used to investigate the jet quenching \cite{M6} of
virtual gluons and thermalization of a strongly-coupled plasma
\cite{Elena}, with a non-zero chemical potential via the
gauge/gravity duality. Thus, Vaidya spacetime can be used to model
interesting physical systems. We would like to point out that
Vaidya spacetime  has also been used to analyze  gravitational
collapse \cite{gc}-\cite{gc12}. In fact, the gravitational
collapse in Vaidya spacetime has been widely studied in different
scenarios \cite{r1}-\cite{r6}. So, the study of gravitational
collapse is an important application to Vaidya spacetime. We would
like to point out that either  black holes or naked singularities
form from such a gravitational collapse. The difference between
these two types of singularities is that black holes are covered
by a boundary known as event horizon beyond which no information
is conveyed to an external observer. But naked singularities are
not covered by any such boundaries. Theoretically, the existence
of naked singularity is important because that would mean that it
is possible to observe the gravitational collapse of an object to
infinite density. The formation of naked singularities in general
relativity have been studied using Vaidya spacetime
\cite{refe1}-\cite{refe2}. As this is an interesting physical
system, and massive gravity is an important modification of
general relativity, we will study the formation of naked
singularities in massive gravity. So, in this paper, we will first
study a time dependent geometry in massive gravity using Vaidya
spacetime. Then we will use this solution  to analyze the
formation of naked singularity in massive gravity.  We will also
study the thermodynamics of such a time dependent solution.

\section{Vaidya Spacetime in Massive gravity}
In this section, we will study a time dependent geometry using
Vaidya spacetime in  massive gravity. The  four dimensional action of massive gravity is given by
 \begin{equation}\label{field}
\mathcal{I}=\int d^{4}x \sqrt{-g}\left[\mathcal{R}+
\mathcal{M}^{2}
\sum_{i}^{4}c_{i}\mathcal{U}_{i}(g,f)+\mathcal{L}_{m} \right],
\end{equation}
where $f$ is a fixed symmetric tensor and is called the reference metric, $c_i$ are constants, $\mathcal{M}$ is the massive gravity
parameter and $\mathcal{U}_i$ are symmetric polynomials of the
eigenvalues of the $d\times d$ matrix
${\mathcal{K}^{\mu}}_{\nu}=\sqrt{g^{\mu\alpha}f_{\alpha\nu}} $
given by
\begin{eqnarray}\label{Ui}
&&\mathcal{U}_{1}=[\mathcal{K}],\nonumber\\
&&\mathcal{U}_{2}=[\mathcal{K}]^2 -[\mathcal{K}^2],\nonumber\\
&&\mathcal{U}_{3}=[\mathcal{K}]^3 -3[\mathcal{K}][\mathcal{K}^2]  + 2[\mathcal{K}^3],\nonumber\\
&&\mathcal{U}_{4}=[\mathcal{K}]^4 -6[\mathcal{K}^2][\mathcal{K}]^2  + 8[\mathcal{K}^3][\mathcal{K}]+3[\mathcal{K}^2]^2
-6[\mathcal{K}^4].
\end{eqnarray}
The square root in $\mathcal{K}$ means
${(\sqrt{A})^{\mu}}_{\nu}{(\sqrt{A})^{\nu}}_{\lambda}={A^{\mu}}_{\lambda}$ and $\mathcal{K}={\mathcal{K}^{\mu}}_{\mu}$.
Then, the equation of motion of this action will be
\begin{equation}\label{field}
G_{\mu\nu}+\mathcal{M}^{2}\chi_{\mu\nu}=T_{\mu\nu},
\end{equation}
where $G_{\mu\nu}$ is the Einstein tensor and $\chi _{\mu \nu }$ is
\begin{eqnarray}\label{chi}
&\chi _{\mu \nu } =&-\frac{c_{1}}{2}\left( \mathcal{U}_{1}g_{\mu \nu }-%
\mathcal{K}_{\mu \nu }\right) -\frac{c_{2}}{2}\left( \mathcal{U}_{2}g_{\mu
\nu }-2\mathcal{U}_{1}\mathcal{K}_{\mu \nu }+2\mathcal{K}_{\mu \nu
}^{2}\right)\nonumber\\
 &&-\frac{c_{3}}{2}(\mathcal{U}_{3}g_{\mu \nu }-3\mathcal{U}_{2}
\mathcal{K}_{\mu \nu } +6\mathcal{U}_{1}\mathcal{K}_{\mu \nu
}^{2}-6\mathcal{K}_{\mu \nu }^{3})\nonumber\\
&&-\frac{c_{4}}{2}(\mathcal{U}_{4}g_{\mu \nu
}-4\mathcal{U}_{3}\mathcal{K}_{\mu \nu
}+12\mathcal{U}_{2}\mathcal{K}_{\mu \nu
}^{2}-24\mathcal{U}_{1}\mathcal{K }_{\mu \nu
}^{3}+24\mathcal{K}_{\mu \nu }^{4})
\end{eqnarray}
Now, we can investigate a Vaidya metric in the context of massive gravity.
We consider the spatial reference metric, in the basis $(t, r, \theta, \phi)$, as follows
 \cite{reference}
\begin{equation}\label{fmetric}
f_{\mu \nu }=diag(0, 0, c^2 h_{ij}),
\end{equation}
where $h_{ij}$ is two dimensional Euclidean metric and $c$ is a
positive constant. The Vaidya metric in the advanced time
coordinate system is given by
\begin{equation}\label{6}
ds^{2} =f(t, r)dt^2 +2dtdr+r^2 d\Omega_{2}^2,
 \end{equation}
 where
\begin{equation}
f(t, r)= -\left(1-\frac{m(t,r)}{r} \right).
\end{equation}
We also consider the supporting total energy-momentum tensor of the
field equation (\ref{field}) in the following form
\begin{equation}
T_{\mu\nu}=T_{\mu\nu}^{(n)}+T_{\mu\nu}^{(m)},
\end{equation}
where $T_{\mu\nu}^{(n)}$ and $T_{\mu\nu}^{(m)}$ are the energy-momentum tensor for the Vaidya null radiation and the energy-momentum
tensor of the perfect fluid
supporting the geometry defined, respectively as
\begin{eqnarray}
&&T_{\mu\nu}^{(n)}=\sigma l_{\mu}l_{\nu},\nonumber\\
&&T_{\mu\nu}^{(m)}=(\rho
+p)(l_{\mu}n_{\nu}+l_{\nu}n_{\mu})+pg_{\mu\nu}.
\end{eqnarray}
where $\sigma$, $\rho$ and $p$ are null radiation density, energy density
and pressure of the perfect fluid, respectively. In this regard, $l_{\mu}$ and $n_{\mu}$
are linearly independent future pointing null vectors as
\begin{equation}
l_{\mu}=\left(1,0,0,0\right),~~~~~\&~~~
n_{\mu}=\left(\frac{1}{2}\left(1-\frac{m(t,r)}{r} \right),-1,0,0
\right).
\end{equation}
satisfying the following conditions
\begin{equation}
l_{\mu}l^{\mu}=n_{\mu}n^{\mu}=0,~~~~\&~~~~l_{\mu}n^{\mu}=-1.
\end{equation}
By this null vectors, the non-vanishing components of the total energy-momentum
tensor will be
\begin{eqnarray}
 T_{00}&=&\sigma+\rho\left(1-\frac{m(t,r)}{r}\right),\nonumber\\
T_{01}&=&-\rho, \nonumber\\
 T_{22}&=&pr^2,\nonumber\\
T_{33}&=&pr^2 sin^2\theta.
\end{eqnarray}
Moreover, using the metric ansatz (\ref{fmetric}), we obtain
\begin{equation}\label{fmetric}
{\mathcal{K}^{\mu}}_{ \nu }=diag\left(0, 0,
\frac{c}{r},\frac{c}{r}\right).
\end{equation} Consequently, we find
\begin{eqnarray}\label{K1}
&&({\mathcal{K}^2)^{\mu}}_{ \nu }={\mathcal{K}^{\mu}}_{ \alpha }{\mathcal{K}^{\alpha}}_{ \nu }=diag\left(0, 0, \frac{c^{2}}{r^{2}},\frac{c^{2}}{r^{2}}\right),\nonumber\\
&&({\mathcal{K}^3)^{\mu}}_{ \nu }={\mathcal{K}^{\mu}}_{ \alpha }{\mathcal{K}^{\alpha}}_{ \beta }{\mathcal{K}^{\beta}}_{ \nu }=diag\left(0, 0, \frac{c^{3}}{r^{3}},\frac{c^{3}}{r^{3}}\right),\nonumber\\
&&({\mathcal{K}^4)^{\mu}}_{ \nu }={\mathcal{K}^{\mu}}_{ \alpha
}{\mathcal{K}^{\alpha}}_{ \beta }{\mathcal{K}^{\beta}}_{ \lambda
}{\mathcal{K}^{\lambda}}_{ \nu }=diag\left(0, 0,
\frac{c^{4}}{r^{4}},\frac{c^{4}}{r^{4}}\right),
\end{eqnarray}
as well as the following required quantities
\begin{eqnarray}\label{K2}
[\mathcal{K}]&=&{\mathcal{K}^{\mu}}_{ \mu }=\frac{2c}{r},\nonumber\\
{[\mathcal{K}^2]}&=&({\mathcal{K}^2)^{\mu}}_{ \mu }=\frac{2c^{2}}{r^{2}},\nonumber\\
{[\mathcal{K}^3]}&=&({\mathcal{K}^3)^{\mu}}_{ \mu }=\frac{2c^{3}}{r^{3}},\nonumber\\
{[\mathcal{K}^4]}&=&({\mathcal{K}^4)^{\mu}}_{ \mu
}=\frac{2c^{4}}{r^{4}}.
\end{eqnarray}
Now, using the equations (\ref{K1}) and (\ref{K2}), we can find the  $\mathcal{U}_i$s in the equation (\ref{Ui}) as
\begin{eqnarray}\label{U}
\mathcal{U}_{1} &=&\frac{2c}{r}, \nonumber\\
\mathcal{U}_{2}&=&\frac{2c^{2}}{r^{2}}, \nonumber\\
\mathcal{U}_{3} &=&0, \nonumber\\
\mathcal{U}_{4} &=&0.
\end{eqnarray}
Using the equations (\ref{fmetric}), (\ref{K1}), (\ref{K2}) and (\ref{U}), we can obtain the non-vanishing components of the massive gravity
term $\chi_{\mu\nu}$ in the field equation
(\ref{field}) as
\begin{eqnarray}
 \chi_{00}&=&\left[\frac{c_{1}c}{r}+ \frac{c_{2}c^2}{r^{2}}\right]\left( 1-\frac{m}{r} \right),\nonumber\\
\chi_{01}&=&\chi_{10}=-\frac{1}{r}\left(c_{1}c+\frac{c_{2}c^2}{r}\right), \nonumber\\
 \chi_{22}&=&-\frac{c_{1}c~r}{2},\nonumber\\
\chi_{33}&=&-\frac{c_{1}~c~r\,sin^{2}\theta}{2}.
\end{eqnarray}
 Then, for the ${00}$ component of the field equation (\ref{field}), we have
\begin{equation}\label{18}
\frac{1}{r^{3}}\left[ r\,\dot m +r m^{\prime} -m\, m^{\prime} \right]
=\sigma+\rho\left(1-\frac{m}{r}\right)-\mathcal{M}^{2}\left[\frac{c_{1}c}{r
}+ \frac{c_{2}c^2}{r^{2} }\right]\left( 1-\frac{m}{r} \right),
\end{equation}
where dot and prime signs denote the derivative with respect to time and
radial coordinates, respectively. For the $01$ and $10$ component of the field equation, we have
\begin{eqnarray}\label{19}
-\frac{m^{\prime}}{r^{2}}=-\rho+\frac{\mathcal{M}^{2}}{r}\left(c_{1}c+\frac{c_{2}c^2}{r}\right).
\end{eqnarray}
Finally, for the $22$ and $33$ component of the field equation, we obtain
\begin{equation}\label{20}
-\frac{1}{2}r
m^{\prime\prime}=pr^2+\frac{\mathcal{M}^{2}c_{1}c~r}{2}.
\end{equation}

\section{Dynamics of the Collapsing System}
In this section, we discuss the  collapsing system dynamics, which is a time dependent system,
in massive gravity. This can be done by finding a solution for the field equations obtained  in the previous section.
We assume that the matter field follows the barotropic equation of state,  which is given by
\begin{equation}\label{21}
p=k\rho,
\end{equation}
where $k$ is the barotropic parameter. Using the equations (\ref{18}),
(\ref{19}), (\ref{20}) and (\ref{21}) we get a solution for $m$ as follows,
\begin{equation}\label{22}
m(t,r)=\frac{r^{1-2k}}{1-2k}f_{1}(t)+f_{2}(t)-\frac{1}{2}\mathcal{M}^{2}cc_{1}r^{2}-\mathcal{M}^{2}c^{2}c_{2}r,~~~~
k\neq 1/2
\end{equation}
where $f_{1}(t)$ and $f_{2}(t)$ are arbitrary functions of $t$
given by $f_{1}(t)=\rho(t,r)r^{2(1+k)}$, and\\
$$\sigma(t,r)=\frac{r^{-(1+2k)}}{1-2k}\dot{f_{1}(t)}+\dot{f_{2}(t)}/r^{2}$$
with $k\neq 1/2$,  where dot represents derivative with respect to t.\\
Therefore, the metric given in the equation (\ref{6}) can be written as
 the generalized Vaidya metric in massive gravity with the metric function\begin{equation}\label{f1}
f(t, r)=-1+\frac{r^{-2k}}{1-2k}f_{1}(t)+\frac{f_{2}(t)}{r}-\frac{1}{2}\mathcal{M}^{2}cc_{1}r-\mathcal{M}^{2}c^{2}c_{2},~~~~
k\neq 1/2.
\end{equation}
Now we will investigate the existence  of naked singularity (NS) in generalized Vaidya
spacetime. This will be done  using the outgoing radial null
geodesics, which will end up in the past at a singularity. So, the
geodesics will terminate  in the central physical singularity
located at $r=0$. It is possible for this singularity to be either
a naked singularity or a black hole (BH).
 Now for a   locally naked singularity, such null
geodesics exist. Furthermore, if the singularity is not a naked singularity, then this system forms a black hole.
Thus, by analyzing the   radial null geodesics that   emerge from the
singularity, we can understand the nature of such a singularity.

{It may be noted that a  singularity can be formed by a catastrophic gravitational
collapse. In general, such a singularity can be either  a naked singularity  or a black hole. However, in  general relativity, such a singularity
formed from a gravitational collapse is always a black hole. This is because of the
 cosmic censorship in general relativity.  So,  in general relativity, the
 singularity is always covered by an event
horizon. However, this need not be the case for a more general theory. In fact,  it is possible  for
inhomogeneous dust cloud  to form  a naked singularity through
a collapse \cite{Eardley1}. It may be noted that some interesting results have been obtained
for fluids whose equations of state is different from
a dust cloud \cite{Joshi1}. So, it is possible to generalize the cosmic censorship in general relativity \cite{Joshi2}.}

{Now, let us consider  $R(t, r )$ as the physical radius at time $t$ of the
shell at $r$. Using the  scaling freedom in this system, we can write  $R(0,r)=r$ at   the starting
time  $t=0$. So, different shells become singular at
different times for  the inhomogeneous
case. It is possible for  future directed radial null
geodesics to come out  of the singularity, with a well defined
tangent at the singularity. So,  $dR/dr$ must tend to a finite limit, as the system  approaches
the past singularity along
these trajectories.}

{ As  the singularity is formed at  $R(t_0,0)=0$, the matter shells are crushed to zero radius at $(t_0,r)=0$.
This  singularity, which is formed at $r=0$,  is called as the  central singularity.}
{Now it is possible for  future directed
non-space like curves to have  their past end points
at this singularity. Such a singularity would then be called as a naked singularity.
So, for such a system,  the outgoing null geodesics
terminate in the past at the central singularity located at $r=0$. This occurs  at
$t=t_0$, and for this point, $R(t_0, 0)=0$. So,  along these geodesics, we obtain
$R\rightarrow 0$ as $r\rightarrow 0$ \cite{Singh1}.}

\begin{figure}[tbp]
\centering
\includegraphics[scale=0.7]{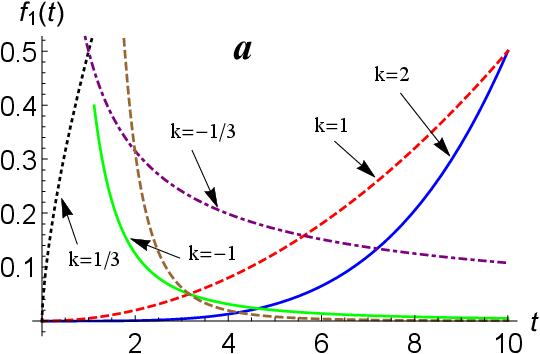}~~~~~~~~~~
\includegraphics[scale=0.7]{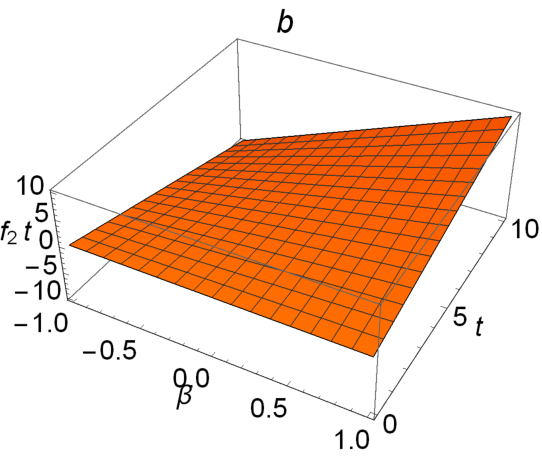}
\caption{\label{fig1ab} Fig (a) shows the variation of
$f_{1}(t)$ with $t$ for different values of $k$. The initial
condition is fixed at $\alpha=0.5$. Fig (b) shows the variation of $f_{2}(t)$
with $t$ and $\beta$.}
\end{figure}

We can write the  equation for the outgoing radial null geodesics using
 the equation (\ref{6}), and  setting $ds^{2}=0$ and
$d\Omega_{2}^{2}=0$. Thus,  we obtain
\begin{equation}
\frac{dt}{dr}=\frac{2}{\left(1-\frac{m(t,r)}{r}\right)}.
\end{equation}
It may be noted  that this system has a singularity at  $r=0,~t=0$. Now, if the function $X$ is given by  $X=\frac{t}{r}$, then
we can study the limiting behavior of $X$ as we
approach the singularity located at $r=0,~t=0$,  along the radial null
geodesic. This limiting value of $X$ will be denoted  by $X_{0}$, and so we can write
\begin{eqnarray}\label{26}
\begin{array}{c}
X_{0}\\\\
{}
\end{array}
\begin{array}{c}
=lim~~ X \\
\begin{tiny}t\rightarrow 0\end{tiny}\\
\begin{tiny}r\rightarrow 0\end{tiny}
\end{array}
\begin{array}{c}
=lim~~ \frac{t}{r} \\
\begin{tiny}t\rightarrow 0\end{tiny}\\
\begin{tiny}r\rightarrow 0\end{tiny}
\end{array}
\begin{array}{c}
=lim~~ \frac{dt}{dr} \\
\begin{tiny}t\rightarrow 0\end{tiny}\\
\begin{tiny}r\rightarrow 0\end{tiny}
\end{array}
\begin{array}{c}
=lim~~ \frac{2}{\left(1-\frac{m(t,r)}{r}\right)}. \\
\begin{tiny}t\rightarrow 0\end{tiny}~~~~~~~~~~~~\\
\begin{tiny}r\rightarrow 0\end{tiny}~~~~~~~~~~~~
 {}
\end{array}
\end{eqnarray}
Using the equations (\ref{22}) and (\ref{26}), we have
\begin{eqnarray}\label{27}
\frac{2}{X_{0}}=
\begin{array}llim\\
\begin{tiny}t\rightarrow 0\end{tiny}\\
\begin{tiny}r\rightarrow 0\end{tiny}
\end{array}\left[1-\frac{r^{-2k}}{1-2k}f_{1}(t)-\frac{f_{2}(t)}{r}+\frac{1}{2}\mathcal{M}^{2}cc_{1}r+\mathcal{M}^{2}c^{2}c_{2}\right], ~~~~ k\neq 1/2.
\end{eqnarray}
Now, choosing $f_{1}(t)=\alpha t^{2k}$~and~$f_{2}(t)=\beta t$, we
obtain an algebraic equation in terms of $X_{0}$ from equation
(\ref{27}) as
\begin{equation}\label{28}
\frac{\alpha}{1-2k}X_{0}^{1+2k}+\beta
X_{0}^{2}-\left(1+\mathcal{M}^{2}c^{2}c_{2}\right)X_{0}+2=0, ~~~~
k\neq 1/2.
\end{equation}
where $\alpha$ and $\beta$ are constants.
It may be noted that naked singularity can also from in  general relativity,
as this equation would have positive roots
even for $\mathcal{M}= 0$.   Such  naked singularities in Vaidya spacetime
have been studied  in  general relativity \cite{refe1}-\cite{refe2}. However,
in this paper, we will analyze the effects of the mass term on the formation
of naked singularity, and observe what effect can such a mass term have on the
formation of naked singularities. As can be observed the above equation,  a non-zero mass term
$\mathcal{M}= 0$
will change the positive roots of this equation.
So, the addition of such a term will change the effect the behavior of this system.
To analyze such a behavior we first observe that
  the expression of
$f_{1}(t)$,   can either be a constant or a non-linear
function of $t$, depending on the EoS, i.e. the cosmological era.
In the early universe ($k\geq 0$), we see that $f_{1}(t)$ grows
with $t$, whereas in the late universe ($k<0$), $f_{1}(t)$ decays
with time. This fact is demonstrated in Fig.\ref{fig1ab}(a). On
the other hand $f_{2}(t)$ is a linear function of $t$. Nature of
$f_{2}(t)$ is demonstrated in Fig.\ref{fig1ab}(b). It can be seen
that these choices of the arbitrary functions $f_{1}(t)$ and
$f_{2}(t)$ are somewhat self-similar in nature. The choice of
$f_{1}(t)$ is driven by the presence of $r^{-2k}$ in the second
term in equation (\ref{27}). Similarly the choice of $f_{2}(t)$ is
based on the presence of the term $1/r$ in the third term in the
equation (\ref{27}). These self-similar choices follow from the
definition of $X_{0}$ given in equation (\ref{26}). We did not
consider non self-similar cases in order to avoid computational
difficulties. Choices other than self-similar ones will leave
residual $r$ or $t$ coordinates in the second and third terms of
equation (\ref{27}). In the limiting condition this will either
result in elimination of terms or creation of mathematically
undefined terms, both of which are undesirable. So this can be
considered a special class of solution given by the self-similar
choice of the functions $f_{1}(t)$ and $f_{2}(t)$.

A black hole will be formed if we obtain only non-positive
solution of this equation. However, if we obtain a positive real
root for this equation, then this system will be described by a
naked singularity. Here it is difficult to find exact solutions
for $X_{0}$ except for some particular values. This is because the
governing equation of the system is a very complicated one. These
exact solutions are given in the following Table 1. It is clear
from the table that certain conditions between the parameters are
required to be satisfied in order to make the solutions positive.
\vspace{4mm}
\begin{center}
\begin{tabular}{|l||l|l|l|}
\hline

~$k$&~~~$Regime$~&~~~~~~~~~~~$Solution 1$~&~~~~~~$Solution 2$~\\
\hline
          &            &                &
\\

~$0$&~~$Dust$~&~$\frac{U-\alpha-\sqrt{\left(U-\alpha\right)^{2}-8\beta}}{2\beta}$~&~$\frac{U-\alpha+\sqrt{\left(U-\alpha\right)^{2}-8\beta}}{2\beta}$~\\
\hline
         &             &                &
\\

~$1$&~$stiff
~fluid$~&~$\frac{\beta}{3\alpha}+\frac{2^{1/3}Y}{3\alpha\left(Z+\sqrt{4Y^{3}+Z^{2}}\right)^{1/3}}-\frac{\left(Z+\sqrt{4Y^{3}+Z^{2}}\right)^{1/3}}{3\times
2^{1/3}\alpha}$~&~~~~~~~~~~~~~~~$-$~~\\
\hline
         &             &                &

\\

$-1/2$&~$DE$~&~$\frac{U-\sqrt{U^{2}-2\beta\left(\alpha+4\right)}}{2\beta}$~&~$\frac{U+\sqrt{U^{2}-2\beta\left(\alpha+4\right)}}{2\beta}$~\\
\hline

\hline
\end{tabular}
\end{center}
\vspace{3mm}
{\bf Table 1:} Exact Values of $X_0$ for specific values of the
EoS parameter $k$ obtained from Eq.(\ref{28}) \vspace{6mm}
where~\\
~~~~~~~~~~~~~~~~~~~~~$U=1+c^{2}c_{2}\mathcal{M}^{2}$,\\\\
~~~~~~~~~~~~~~~~~~~~~$Y=3\alpha+3c^{2}c_{2}\mathcal{M}^{2}\alpha-\beta^{2}$,\\\\
~~~~~~~~~~~~~~~~~~~~~$Z=-54\alpha^{2}+9\alpha\beta+9c^{2}c_{2}\mathcal{M}^{2}\alpha\beta-2\beta^{3}$.\\\\
\begin{figure}[tbp]
\centering
\includegraphics[scale=0.7]{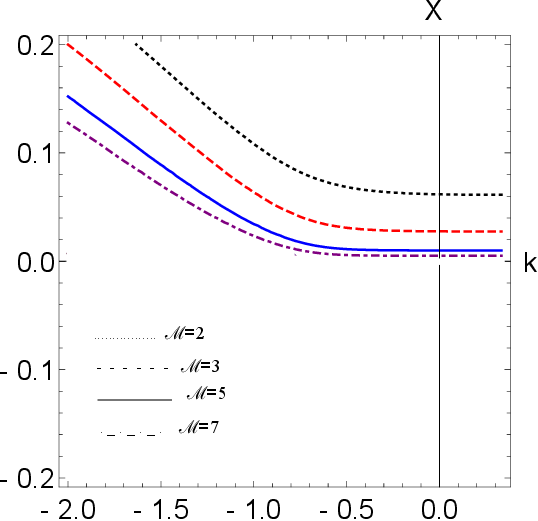}
\includegraphics[scale=0.7]{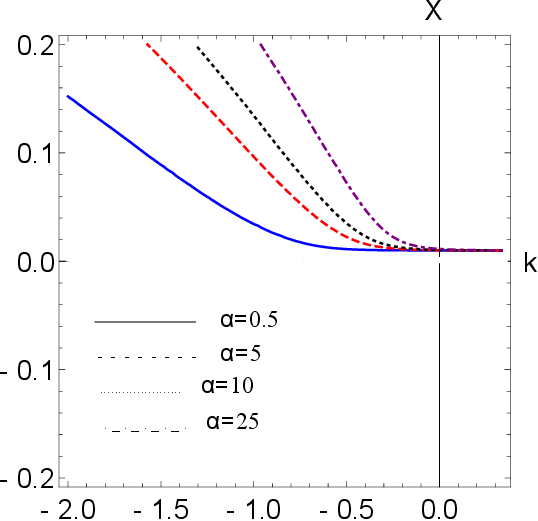}
\caption{\label{fig2ab} Figs (a) and (b) show the
variation of $X_{0}$ with $k$ for different values of
$\mathcal{M}$ and $\alpha$ respectively. In Fig(a) the
initial conditions are fixed at $\alpha=0.5$, $\beta=2$, $c=2$,
$c_{2}=2$. In Fig.(b) the initial conditions are fixed at
$\mathcal{M}=5$, $\beta=2$, $c=2$, $c_{2}=2$.}
\end{figure}
We now analyze the results furnished in the Table 1.\\

\textit{\textbf{Case 1: k=0}}\\\\
This corresponds to the pressureless dust regime of the universe.
For the solutions to be real and finite, we must have $\beta\neq
0$, $\left(U-\alpha\right)^{2}\geq 8\beta$.
\\
For positivity of the first solution we have for
$\beta>0$,~~$U-\alpha\geq\sqrt{\left(U-\alpha\right)^{2}-8\beta}\Longrightarrow
\beta\geq 0$. But since $\beta\neq 0$, we must restrict ourselves
to $\beta>0$, which is in agreement with our assumption. So the
first solution is positive for any positive $\beta$. For
$\beta<0$, we have for positive solution
$U-\alpha\leq\sqrt{\left(U-\alpha\right)^{2}-8\beta}\Longrightarrow
\beta\leq 0$. But since $\beta\neq 0$, we must have $\beta<0$
which is in agreement with our assumption. So, for $\beta<0$,
the solution 1 is always positive and represents a NS. Hence, for any
non-zero real $\beta$ the first solution represents an NS.
\\
For solution 2 to be positive we must have for $\beta>0$,
$U-\alpha\geq
-\sqrt{\left(U-\alpha\right)^{2}-8\beta}\Longrightarrow\beta\geq
0$. This gives $\beta>0$, just like the previous case. Hence, the
solution is positive. Similarly for $\beta<0$ case also we get
positive solution. Thus, this solution also represents NS.
Therefore for $k=0$, we get NS as the end state of collapse.\\\\

\textit{\textbf{Case 2: k=1}}\\\\
This corresponds to the early stiff fluid era of our universe.
Here, the situation is much more chaotic mathematically. We get
only one solution, which turns out to be a relatively complicated
one. Now physically speaking, in the early universe, due to big
bang extreme amount of chaos is expected. Moreover, there are
quantum fluctuations, so mathematically the scenario is justified.
Here, in order to have a positive solution we should have
$\frac{\beta}{3\alpha}+\frac{2^{1/3}Y}{3\alpha\left(Z+\sqrt{4Y^{3}+Z^{2}}\right)^{1/3}}\geq
\frac{\left(Z+\sqrt{4Y^{3}+Z^{2}}\right)^{1/3}}{3\times
2^{1/3}\alpha}$. Moreover for negative $Y$, $Z^{2}\geq 4Y^{3}$ for
the solution to be real.\\\\

\textit{\textbf{Case 3: k=-1/2}}\\\\
This represents the dark energy era corresponding to the late time
accelerated expanding universe. Here, the solution to be real and
finite we have, $\beta\neq 0$, $U^{2}\geq
2\beta\left(\alpha+4\right)$. For positive solution, we must have
$\beta\left(\alpha+4\right)\geq 0$ in which for $\beta>0$ $\Longrightarrow$
$\alpha\geq -4$. For $\beta<0$, we should have
$\beta\left(\alpha+4\right)\leq 0$ $\Longrightarrow$ $\alpha\leq
-4$ in order to get a positive solution.\\\\

~~~~~~~~~~~~~~~\textit{\textbf{Numerical Solutions of $X_0$ and their interpretations}}\\\\
In order to understand the dynamics of collapse, we need to have a
knowledge of $X_{0}$ not at discrete points of $k$, but throughout
the cosmologically meaningful region $k<1$, i.e., from early to
late universe. To achieve this we proceed to obtain numerical
solutions of $X_{0}$, by assigning different initial conditions to
the parameters describing this system. To visualize these
solutions we obtain contours for $k-X_{0}$ for different numerical
values of the involved parameters.

It may be noted from  the plots that the trajectories run across
the positive range of $X_{0}$ thus confirming the formation of NS.
In Fig.\ref{fig2ab}(a), we can observe the dependence of $X_0$ on
the EoS parameter $k$ for different values of the massive gravity
parameter $\mathcal{M}$. We see that an increase in the value of
$\mathcal{M}$ decreases the tendency of formation of NS. Hence we
observe that the dynamics of the system gets deformed by the
addition of graviton mass to this system. In Fig.\ref{fig2ab}(b),
the $k-X_{0}$ trajectories for different values of $\alpha$ are
obtained. Here also different values of $\alpha$ deform the
dynamics of this system. Greater the value of $\alpha$, greater is
the tendency to form NS.\\\\

\begin{figure}[tbp]
\centering
\includegraphics[scale=0.7]{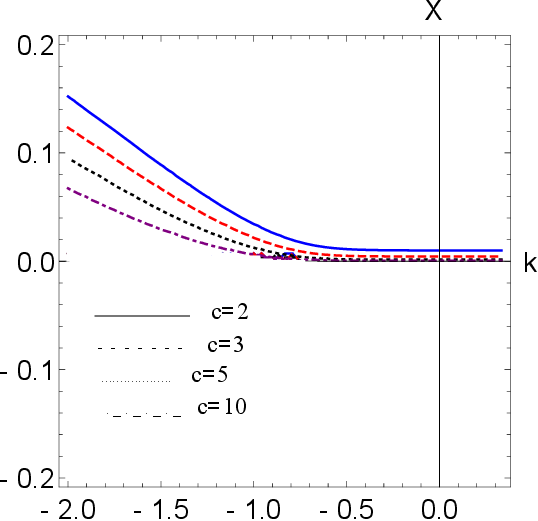}~~~~
\includegraphics[scale=0.7]{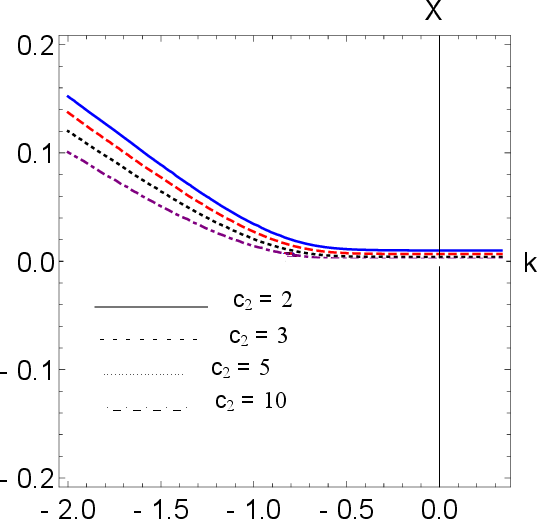}
\caption{\label{fig3ab} Figs (a) and (b) show the
variation of $X_{0}$ with $k$ for different values of $c$ and
$c_{2}$ respectively.In Fig.3(a) the initial conditions are
fixed at $\alpha=0.5$, $\beta=2$, $c_{2}=2$, $\mathcal{M}=5$.
In Fig.3(b) the initial conditions are taken as $\alpha=0.5$,
$\beta=2$, $c=2$, $\mathcal{M}=5$.}
\end{figure}

In Fig.\ref{fig3ab}(a), we observe the effect of $c$ on the
collapsing system. It is observed that greater the value of $c$,
lesser is the tendency to form NS. In Fig.\ref{fig3ab}(b), we can
observe the effect of $c_{2}$ on the system. Here also we see that
an increase in $c_{2}$ decreases the possibility of NS.

\begin{figure}[tbp]
\centering
\includegraphics[scale=0.7]{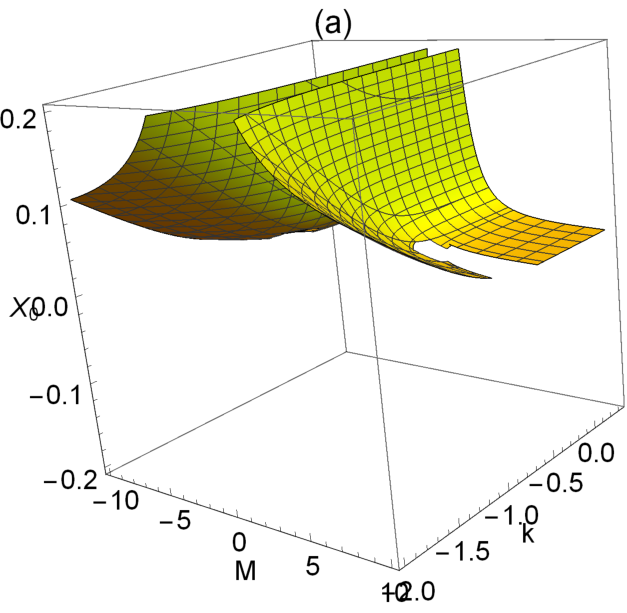}
\includegraphics[scale=0.7]{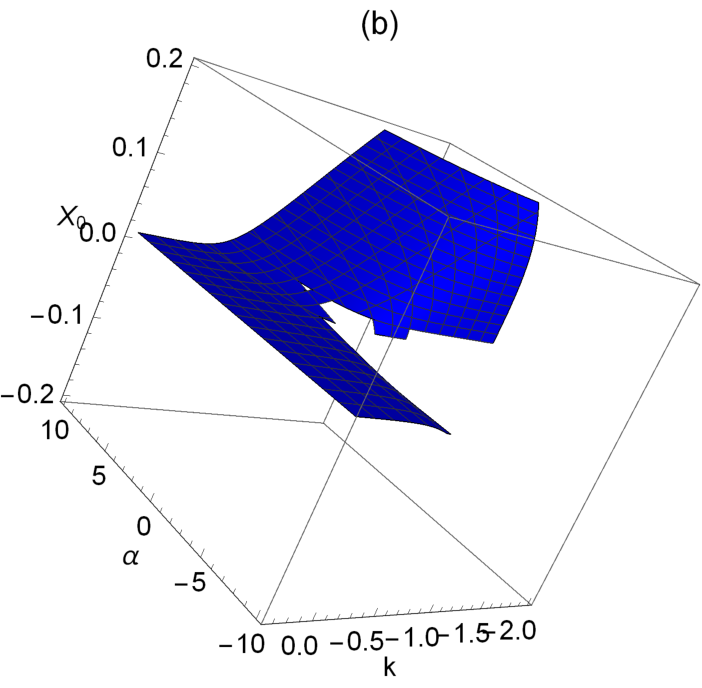}
\caption{\label{fig4ab} Fig (a) shows the variation of
$X_{0}$ with $k$ and $\mathcal{M}$. The initial conditions are
fixed at $\alpha=0.5, \beta=2$, $c=2$, $c_{2}=2$. Fig (b) shows the variation of $X_{0}$ with $k$ and $\alpha$. The initial conditions are taken as
$\mathcal{M}=5, \beta=2$, $c=2$, $c_{2}=2$.}
\end{figure}

\begin{figure}[tbp]
\centering
\includegraphics[scale=0.7]{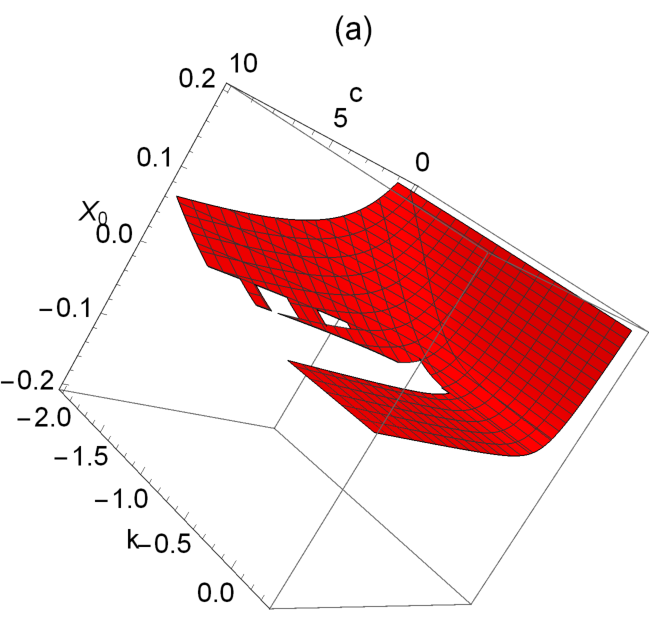}
\includegraphics[scale=0.7]{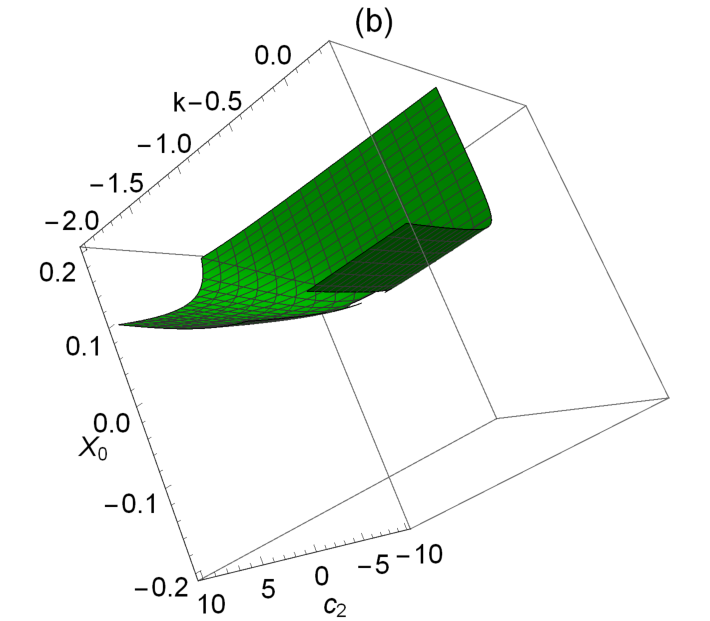}
\caption{\label{fig5ab} Fig (a) shows the variation of
$X_{0}$ with $k$ and $c$. The initial conditions are fixed at
$\alpha=0.5,\beta=2$, $\mathcal{M}=5$, $c_{2}=2$. Fig (b) shows the variation of $X_{0}$ with $k$ and $c_{2}$. The initial conditions are taken
as $\alpha=0.5, \beta=2$, $\mathcal{M}=5$, $c=2$.}
\end{figure}

\begin{figure}[tbp]
\centering
\includegraphics[scale=0.7]{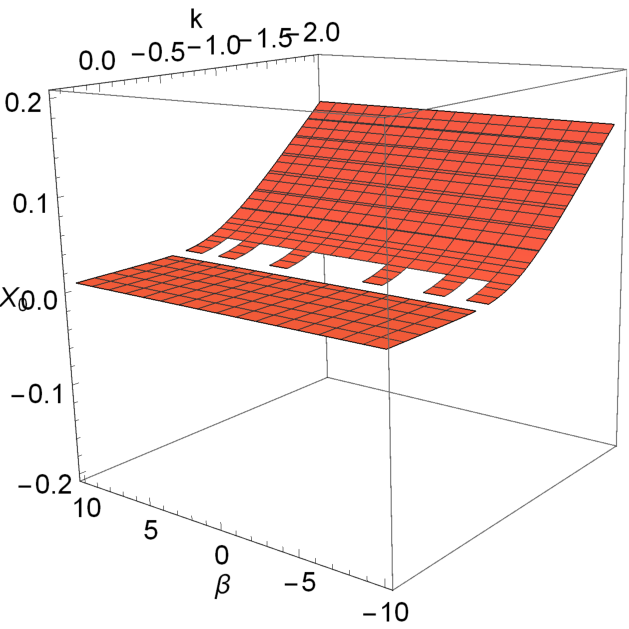}
\caption{\label{fig6a} Fig 6 shows the variation of $X_{0}$
with $k$ and $\beta$. The initial conditions are fixed at
$\mathcal{M}=5$, $c=2$, $c_{2}=2$, $\alpha=0.5$.}
\end{figure}

In Figs.\ref{fig4ab}, \ref{fig5ab} and \ref{fig6a}, 3D-plots are
obtained to get a more comprehensive view of the dynamics of the
collapse. In all the figures the resulting surfaces entirely lie
in the positive half-space of $X_{0}$, thus showing the presence
of NS. In Fig.\ref{fig4ab}(a), the variation of $k-\mathcal{M}$
surfaces are obtained against $X_{0}$. We see that in the dark
energy regime $(k<-1/3)$ the surface pushes towards the positive
direction of $X_{0}$, accompanied with a decrease in
$\mathcal{M}$, thus showing an increased tendency to form NS. In
Fig.\ref{fig4ab}(b), $k-\alpha$ surfaces are obtained against
$X_{0}$. Here also in the dark energy regime, there is an
increased tendency of NS accompanied by an increase in $\alpha$.
In Fig.\ref{fig5ab}(a), $k-c$ surface is obtained against $X_{0}$.
Here accompanied by a decrease in the value of $c$, we witness an
increased tendency of NS in the dark energy regime. In
Fig.\ref{fig5ab}(b), $k-c_{2}$ surface is obtained against
$X_{0}$. The results obtained are same as that of
Fig.\ref{fig5ab}(a). In Fig.\ref{fig6a}, $k-\beta$ surface is
obtained against $X_{0}$. We see that the $k-\beta$ surface is
parallel to the $\beta$ axis. This shows that the system is not
deformed by $\beta$ and hence the collapse dynamics does not
depend on it. This is an important result. Finally just like the
previous cases here also the surface gets lifted in the dark
energy regime towards the positive direction of $X_{0}$ axis.\\\\

 \textit{\textbf{Strength of Singularity}}\\\\
 It is important to know about the destructive capacity of a singularity, and this is measured using the 
 concept of  strength of singularity. The strength of singularity is related to the extension 
 of spacetime through the singularity. Now this can be quantified  using the Tipler's  formalism \cite{Tipler}-\cite{Maharaj}. 
 Now using the  Tipler's formalism  \cite{Tipler}-\cite{Maharaj}, the 
condition for a singularity to be strong is given by,
\begin{eqnarray}
\begin{array}{c}
S=lim~~ \tau^{2}\psi \\
\begin{tiny}\tau\rightarrow 0\end{tiny}\\
\end{array}
\begin{array}{c}
=lim~~ \tau^{2}R_{\mu\nu}K^{\mu}K^{\nu}>0 \\
\begin{tiny}\tau\rightarrow 0\end{tiny}\\
\end{array}
\end{eqnarray}
where $R_{\mu\nu}$ is the Ricci tensor. Here $\psi$ is a scalar
given by $\psi=R_{\mu\nu}K^{\mu}K^{\nu}$,  
$K^{\mu}=\frac{dx^{\mu}}{d\tau}$ is the tangent to the non-spacelike geodesics at the singularity,  
and $\tau$ is the affine
parameter.  It has been demonstrated  that \cite{Maharaj}, 
\begin{eqnarray}\label{maha}
\begin{array}{c}
S=lim~~ \tau^{2}\psi \\
\begin{tiny}\tau\rightarrow 0\end{tiny}\\
\end{array}
\begin{array}{c}
=\frac{1}{4}X_{0}^{2}\left(2\dot{m_{0}}\right) \\
\begin{tiny}~\end{tiny}\\
\end{array}
\end{eqnarray}
where $m_0$ is given by 
\begin{eqnarray}
\begin{array}{c}
m_{0}=lim~~ m(t,r) \\
\begin{tiny}t\rightarrow 0\end{tiny}\\
\begin{tiny}r\rightarrow 0. \end{tiny}
\end{array}
\end{eqnarray}
Furthermore, it is also possible to write 
\begin{eqnarray}\label{massd}
\begin{array}{c}
\dot{m_{0}}=lim~~ \frac{\partial}{\partial~t}\left(m(t,r)\right) \\
\begin{tiny}t\rightarrow 0\end{tiny}\\
\begin{tiny}r\rightarrow 0.\end{tiny}
\end{array}
\end{eqnarray}

\begin{figure}[tbp]
\centering
\includegraphics[scale=0.7]{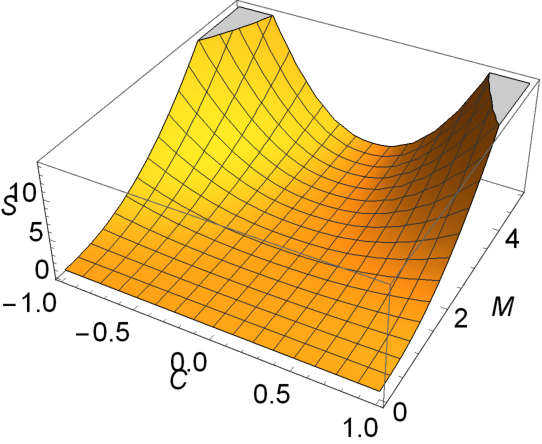}
\caption{\label{fig7} The figure shows the variation of $S$ with
$c$ and $\mathcal{M}$. The initial conditions are fixed at
$\beta=2$, $c_{2}=2$, $\alpha=0.5$.}
\end{figure}

Using Eq. (\ref{22}) in the above relation (\ref{maha}), we obtain 
\begin{eqnarray}\label{strength}
\begin{array}{c}
S=lim~~ \tau^{2}\psi \\
\begin{tiny}\tau\rightarrow 0\end{tiny}\\
\end{array}
\begin{array}{c}
=\frac{1}{4}X_{0}^{2}\left[\frac{2k\alpha}{1-2k}X_{0}^{2k-1}+\beta\right] \\
\begin{tiny}~\end{tiny}\\
\end{array}
\end{eqnarray}
It may be noted that it has been demonstrated that  $X_{0}$ is related to
the limiting values of mass as \cite{Maharaj}
\begin{equation}\label{xmass}
X_{0}=\frac{2}{1-2m_{0}'-2\dot{m_{0}}X_{0}}
\end{equation}
where $m_{0}'$ is given by 
\begin{eqnarray}\label{dashedmass}
\begin{array}{c}
m_{0}'=lim~~ \frac{\partial}{\partial~r}\left(m(t,r)\right) \\
\begin{tiny}t\rightarrow 0\end{tiny}\\
\begin{tiny}r\rightarrow 0. \end{tiny}
\end{array}
\end{eqnarray}
Here $\dot{m_{0}}$ is given by the Eq. (\ref{massd}). Now using
Eqs. (\ref{22}), (\ref{massd}) and (\ref{dashedmass}) in
Eq. (\ref{xmass}), we obtain 
\begin{equation}\label{xf}
\frac{2\alpha}{1-2k}X_{0}^{2k+1}+2\beta
X_{0}^{2}-\left(1+2\mathcal{M}^{2}c^{2}c_{2}\right)X_{0}-2=0
\end{equation}
In a particular case, if $k=-1/2$ (dark energy) is considered, we
obtain a solution for Eq. (\ref{xf}) as
\begin{equation}
X_{0}=\frac{0.25}{\beta}\left(1+2c^{2}c_{2}\mathcal{M}^{2}+\sqrt{8\beta\left(2-\alpha\right)+\left(1+2c^{2}c_{2}\mathcal{M}^{2}\right)^{2}}\right)
\end{equation}
which is positive when~ $1+2c^{2}c_{2}\mathcal{M}^{2} \geq
2\sqrt{2\beta\left(\alpha-2\right)}$ and $\alpha>2$ for $\beta>0$, 
provided   $c_{2}>0$. The existence of such
a positive root signifies that the singularity is naked. Using the
above value of $X_{0}$ in the Eq. (\ref{strength}), we get
\begin{eqnarray}\label{strength}
\begin{array}{c}
S=lim~~ \tau^{2}\psi \\
\begin{tiny}\tau\rightarrow 0\end{tiny}\\
\end{array}
\begin{array}{c}
=\frac{0.03125}{\beta}u^{2}\left(1-\frac{8\alpha \beta}{u^{2}}\right) \\
\begin{tiny}~\end{tiny}\\
\end{array}
\end{eqnarray}
where we have 
$u=1+2c^{2}c_{2}\mathcal{M}^{2}+\sqrt{8\beta\left(2-\alpha\right)+\left(1+2c^{2}c_{2}\mathcal{M}^{2}\right)^{2}}$.\\
Here, we can write 
\begin{eqnarray}\label{strength}
\begin{array}{c}
S=lim~~ \tau^{2}\psi >0 \\
\begin{tiny}\tau\rightarrow 0\end{tiny}\\
\end{array}
\end{eqnarray}
for $8\alpha\beta < u^{2}$ for $\beta>0$. This is the condition
for a strong naked singularity. A $3D$ plot for $S$ is
shown in Fig.\ref{fig7} for a particular scenario. The plot shows
that the surface lies in the positive region thus giving a strong
naked singularity.

\section{Thermodynamics}
\begin{figure}[tbp]
\centering
\includegraphics[scale=0.35]{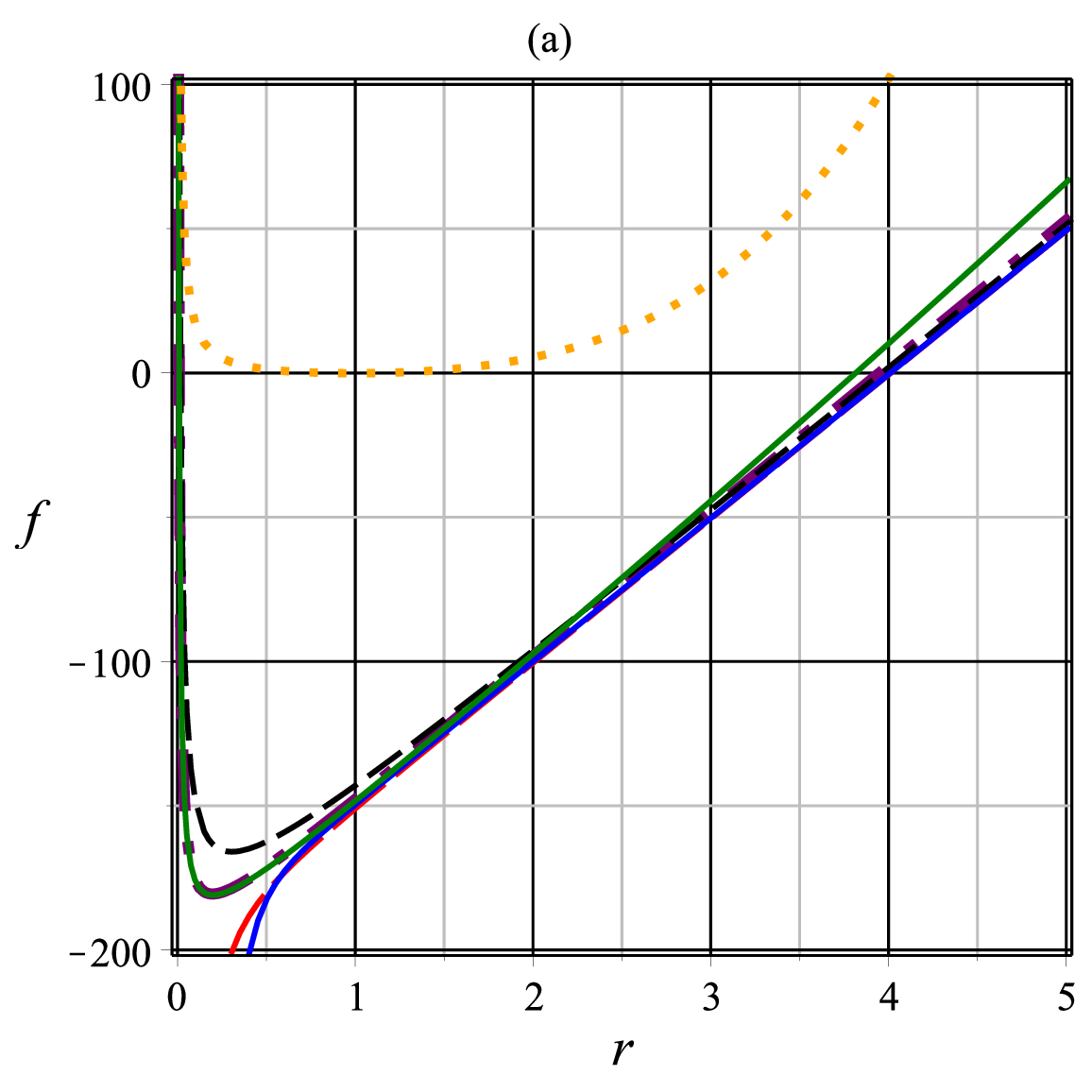}
\includegraphics[scale=0.35]{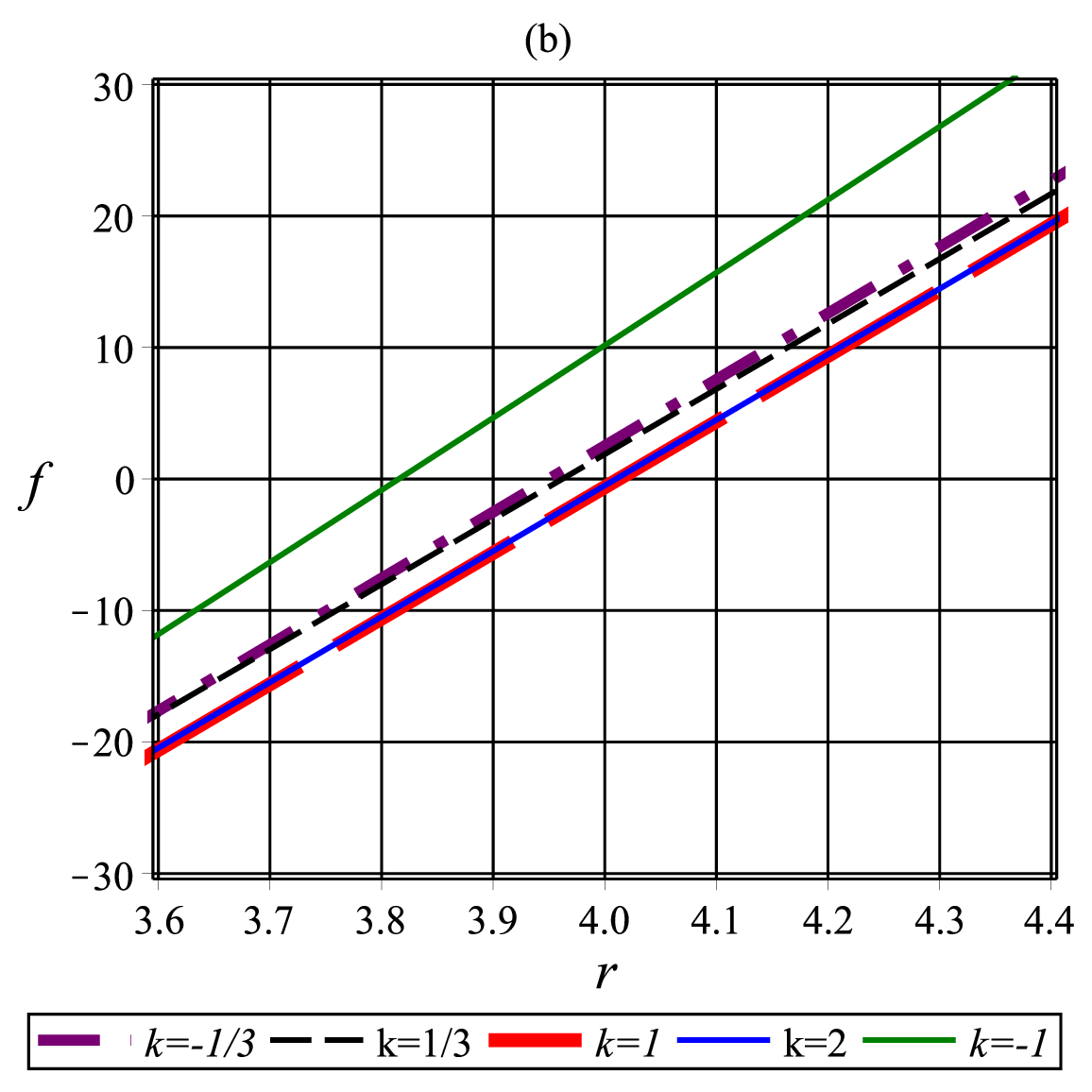}
\caption{\label{fig8}Horizon structure of the
generalized Vaidya spacetime in massive gravity.
It is set $f_{1}(t)=f_{2}(t)=c=c_{1}=c_{2}=2$ and $\mathcal{M}=5$ and the dotted
orange line in plot (a) represents $k=-2$ and $\mathcal{M}=0.5$.
Plot (b) shows the zoomed range of outer horizon given by the plot
(a).}
\end{figure}

In this section, we would like to study the thermodynamics of  generalized Vaidya spacetime in massive gravity.
The thermalization temperature, for such a spacetime, is given by the following relation \cite{Elena},
\begin{equation}\label{T1}
T=\frac{1}{4\pi}\frac{d}{dr}f(t,r)|_{r=r_{h}},
\end{equation}
where $r_{h}$ is the event horizon obtained from the following relation (see the equation (\ref{f1})),
\begin{equation}\label{T2}
-1+\frac{r^{-2k}}{1-2k}f_{1}(t)+\frac{f_{2}(t)}{r}-\frac{1}{2}\mathcal{M}^{2}cc_{1}r-\mathcal{M}^{2}c^{2}c_{2}=0.
\end{equation}
Real positive root of the above equation gives the event horizon
radius. In Fig.\ref{fig8}, we can see the typical behavior of
$f(t,r)$ in terms of $r$. We show that it is possible to have two radii at
which $f(t,r)=0$, and the bigger one (solid red) shows the event
horizon radius (about $r=4$ of Fig.\ref{fig8}). We can also see
that the increasing value of $k$ increases the value of outer
event horizon radius. In the case of $k>0.5$ we can see only one zero (see red and blue lines). Also, we can see extremal case with $k=-2$
and $\mathcal{M}=0.287$. We should note that having one or two
horizons is a function of the massive parameter $\mathcal{M}$.

\begin{figure}[tbp]
\centering
\includegraphics[scale=0.35]{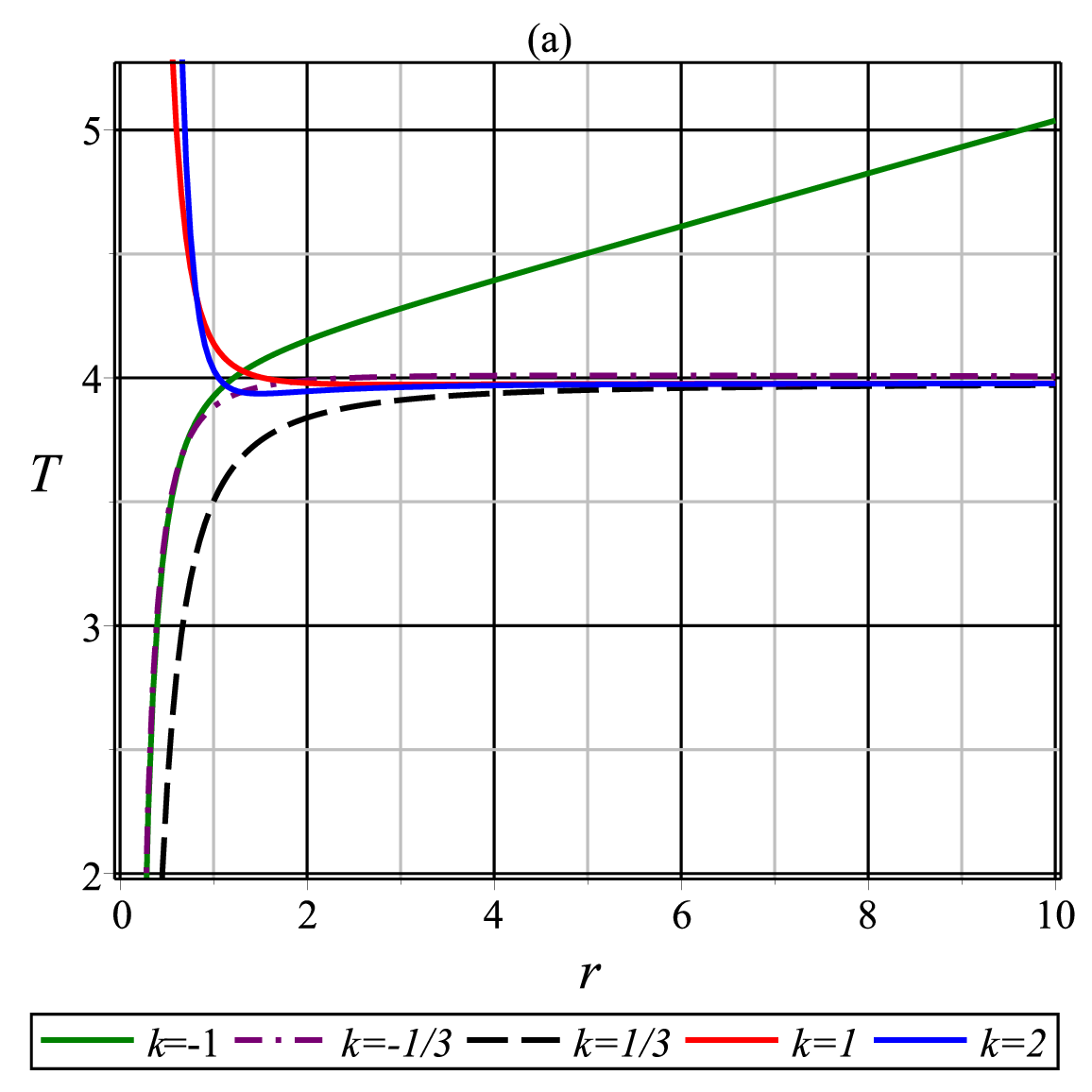}
\includegraphics[scale=0.35]{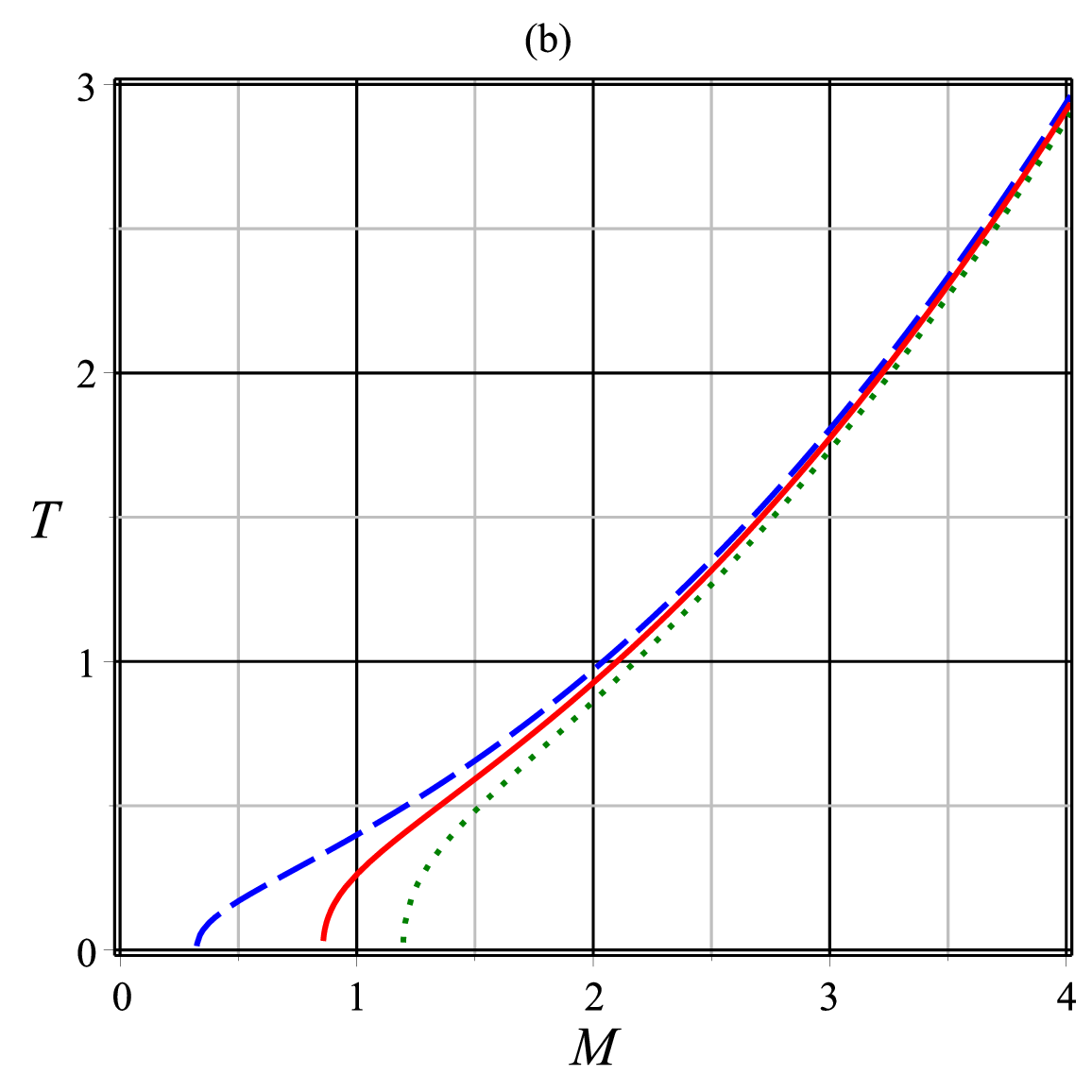}
\caption{\label{fig9}Typical behavior of the
temperature. (a) in terms of radius; (b) in terms of $\mathcal{M}$
for $f_{1}(t)=2$ and $f_{2}(t)=1, 5, 9$ respectively denoted by
dashed blue, solid red and dotted green lines. $c=c_{1}=c_{2}=2$,
and $\mathcal{M}=5$, and $k=-1$.}
\end{figure}

In the special case of $k=-1$, one can obtain real positive root as,
\begin{equation}\label{T3}
r_{h}=\frac{1}{2f_{1}}\left(Y^{\frac{1}{3}}+\frac{c_{1}^{2}c^{2}\mathcal{M}^{4}+4f_{1}(c_{2}c^{2}\mathcal{M}^{2}+1)}{Y^{\frac{1}{3}}}-c_{1}c\mathcal{M}^{2}\right),
\end{equation}
where we defined,
\begin{equation}\label{T4}
Y=2\sqrt {3}f_{1}\sqrt {M}-12f_{2}{f_{1}}^{2}-6\left( c_{2}{c}^{2}{\mathcal{M}}^{2}+1 \right) {\mathcal{M}}^{2}c_{1}cf_{1}-{\mathcal{M}}^{6}{c_{1}}
^{3}{c}^{3},
\end{equation}
where
\begin{eqnarray}\label{T5}
M&=&-3{c_{1}}^{2}{c_{2}}^{2}{c}^{6}{\mathcal{M}}^{8}+ \left( -16\,f_{1}{c_{2}}^{3}
{c}^{6}+6f_{2}{c_{1}}^{3}{c}^{3}-6{c_{1}}^{2}c_{2}{c}^{4} \right) {\mathcal{M}}^{6}\nonumber\\
&+& \left(36\,f_{1}f_{2}c_{1}c_{2}{c}^{3}-48\,f_{1}{c_{2}}^{2}{c}^{4}-3{c_{1}}^{2}{c}^{2} \right) {\mathcal{M}}^{4}\nonumber\\
&+&36\,f_{1}c \left( f_{2}c_{1}-\frac{4}{3}c_{2}c \right) {\mathcal{M}}^{2}+36{f_{1}}^{2}{f_{2}}^{2}-16\,f_{1}.
\end{eqnarray}
By using the equation (\ref{T1}) one can obtain,
\begin{equation}\label{T6}
T=\frac{1}{4\pi}\left(\frac{{\mathcal{M}}^{2}}{2c_{1}c}-\frac{2kf_{1}}{1-2k}r_{h}^{-2k-1}-f_{2}r_{h}^{-2}\right).
\end{equation}
In the plots of the Fig.\ref{fig9} we can see typical behavior of the temperature for some values of $k$ in terms of $r$ (a) and in terms of $\mathcal{M}$ (b). Fig.\ref{fig9} (a) plotted in terms of $r$, however from the Figs.\ref{fig8} we show that selected value of parameters yields $r_{h}\approx4$. Instead of small radius, in the special case of $k=-1$, it is approximately linear function of radius. We can see that for the case of $k<0.5$, temperature is increasing function of $r$ to yields a constant for the large radius. Situation is vise for the cases of $k>0.5$ (see red and blue solid lines).\\
Then, by using the relation (\ref{T3}) one can obtain temperature in terms of $\mathcal{M}$ in which we can find it as increasing function of $\mathcal{M}$. Also, we can see that increasing $\beta$ ($f_{2}$) decreases the temperature.

\begin{figure}[tbp]
\centering
\includegraphics[scale=0.35]{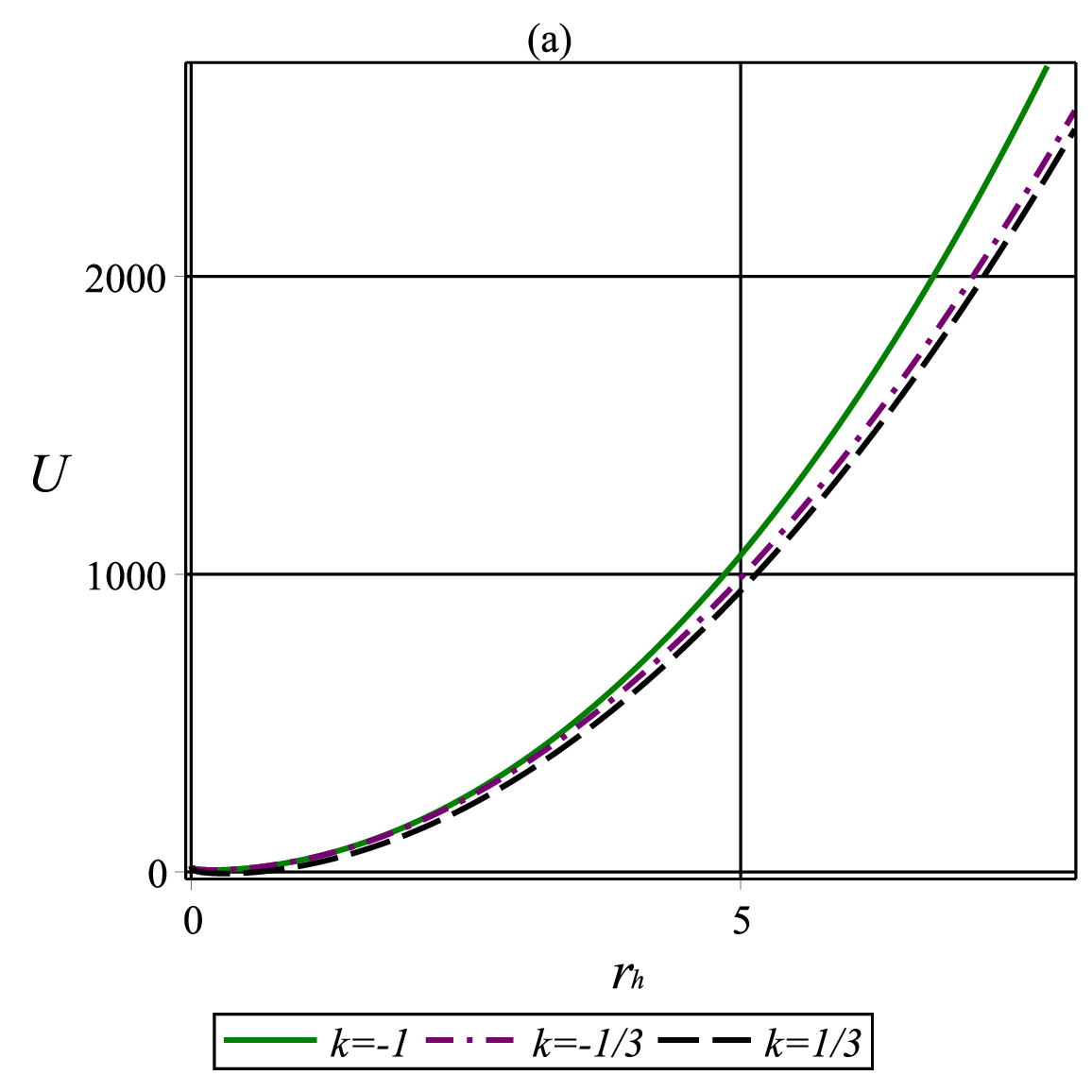}
\includegraphics[scale=0.35]{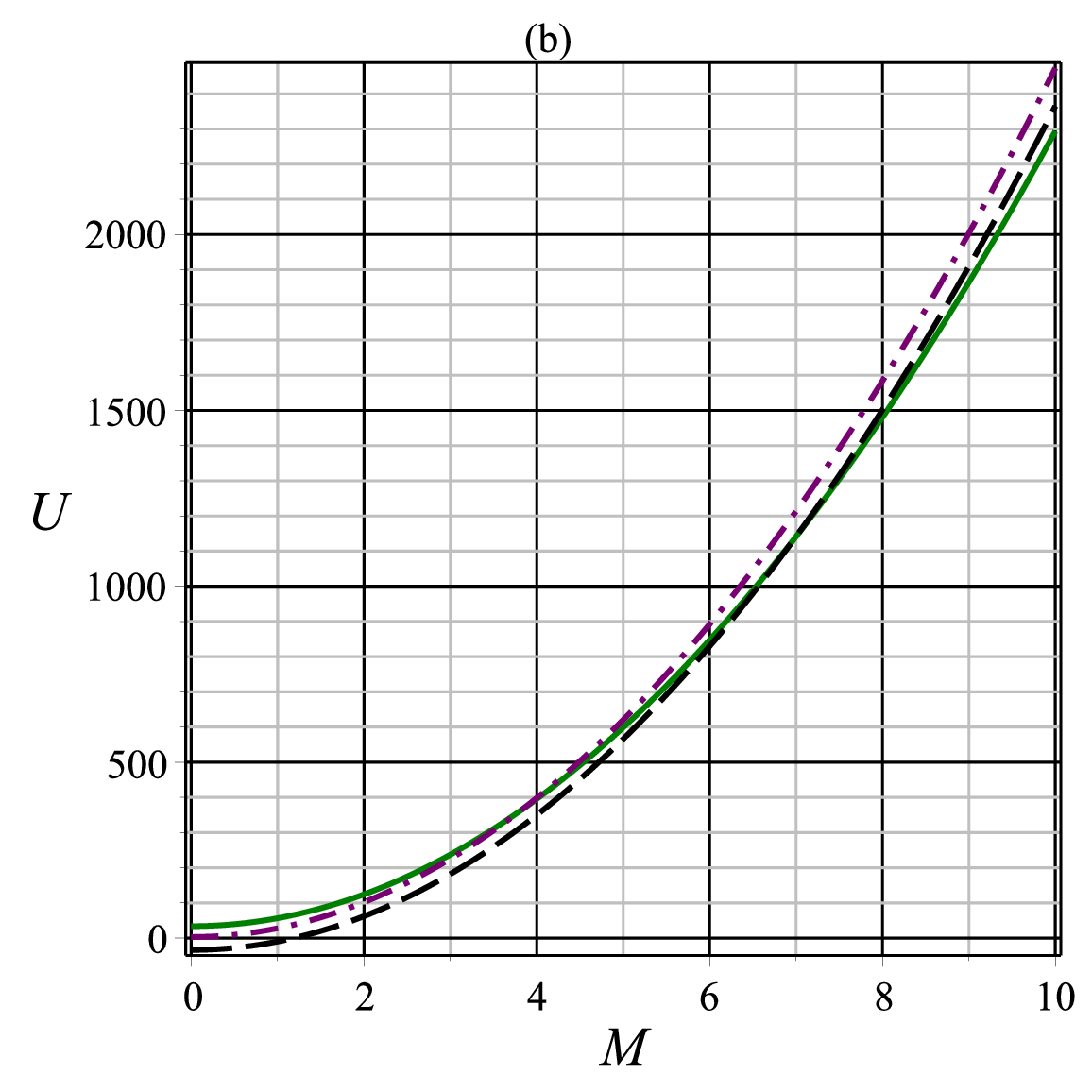}
\caption{\label{fig10}Typical behavior of the
internal energy in terms of (a) radius with $\mathcal{M}=5$ and
(b) mass parameter $\mathcal{M}$ with $r_{h}\approx4$ for
$f_{1}(t)=f_{2}(t)=c=c_{1}=c_{2}=2$.}
\end{figure}

Now, we can write entropy as,
\begin{equation}\label{TS}
S=\pi^{2} r_{h}^{2},
\end{equation}
where we used $\pi G=1$. Hence, we can use the following relation to calculate total energy,
\begin{equation}\label{TU}
U=\int{TdS},
\end{equation}
which yields to the following expression,
\begin{equation}\label{TU-1}
U=-\frac{1}{2}\,\pi \,f_{2}\ln  \left( r_{h} \right) +\frac{1}{8}{\mathcal{M}}^{2}\pi \,c_{1}c{r_{h}}^{2}-\frac{\pi k f_{1}}{4k^{2}-4k+1}r_{h}^{1-2k}.
\end{equation}
In the Fig.\ref{fig10} (a) we can see typical behavior of $U$ for some values of $k$ and find that value of $k$ reduces value of the internal energy. Also, in the Fig.\ref{fig10} (b) we can see that internal energy is increasing function of mass parameter. Internal energy may be used to obtain Helmholtz free energy.

\begin{figure}[tbp]
\centering
\includegraphics[scale=0.35]{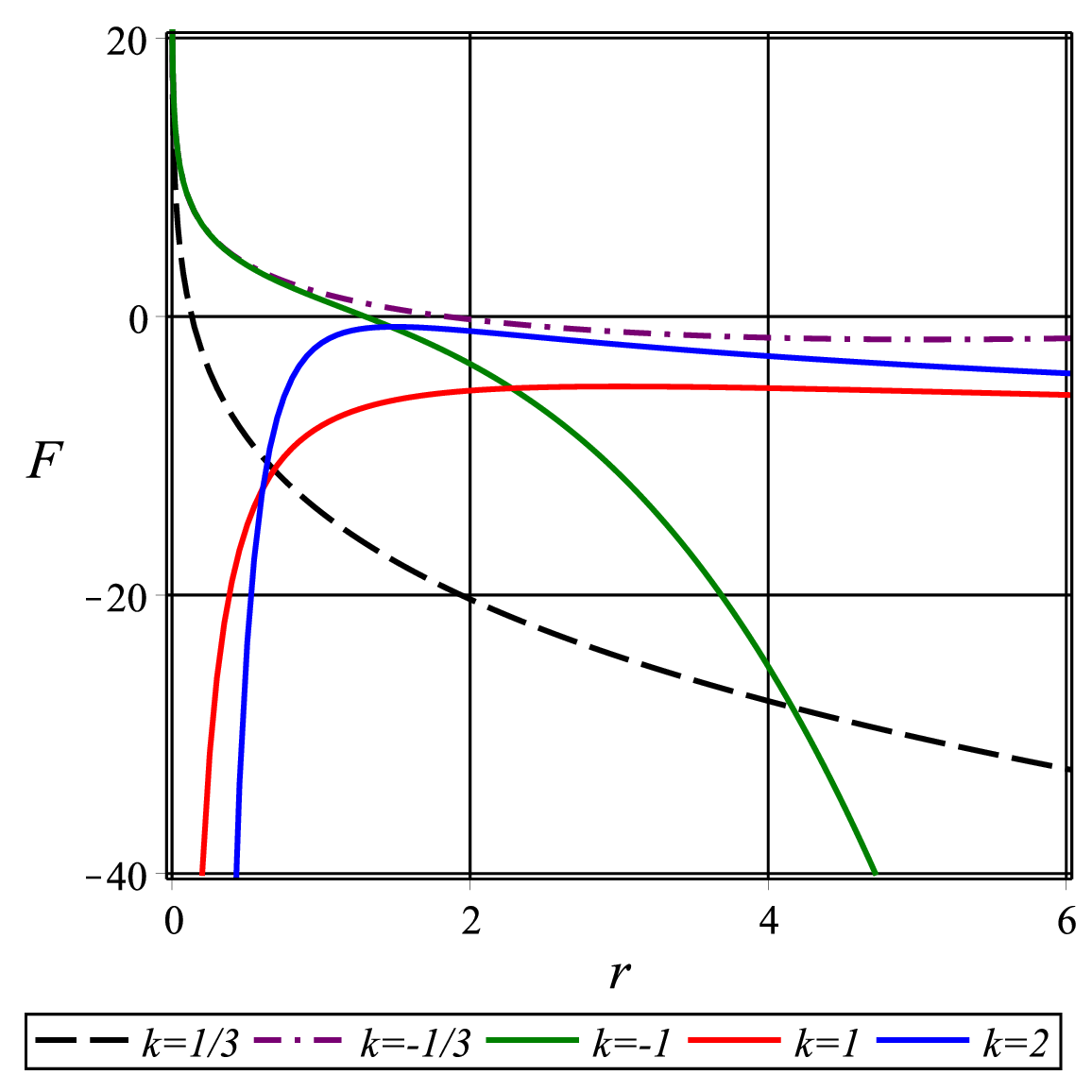}
\caption{\label{fig11} Typical behavior of the
Helmholtz free energy in terms of radius for
$f_{1}(t)=f_{2}(t)=c=c_{1}=c_{2}=2$.}
\end{figure}

Helmholtz free energy can be obtained via the following relation,
\begin{equation}\label{F}
F=U-TS,
\end{equation}
which yields to the following expression,
\begin{equation}\label{F2}
F=\frac{kf_{1}r_{h}^{1-2k}+\sqrt{k-\frac{1}{2}}\left[kf_{1}r_{h}^{1-2k}+2f_{2}(k-\frac{1}{2})(\ln{r_{h}}-\frac{1}{2})\right]}{-\frac{4}{\pi}\sqrt{k-\frac{1}{2}}}.
\end{equation}
It is interesting to note that Helmholtz free energy is independent of $\mathcal{M}$.
In the plots of the Fig.\ref{fig11}, we can see typical behavior of the Helmholtz free energy in terms of radius for various values of $k$.\\\\

\begin{figure}[tbp]
\centering
\includegraphics[scale=0.35]{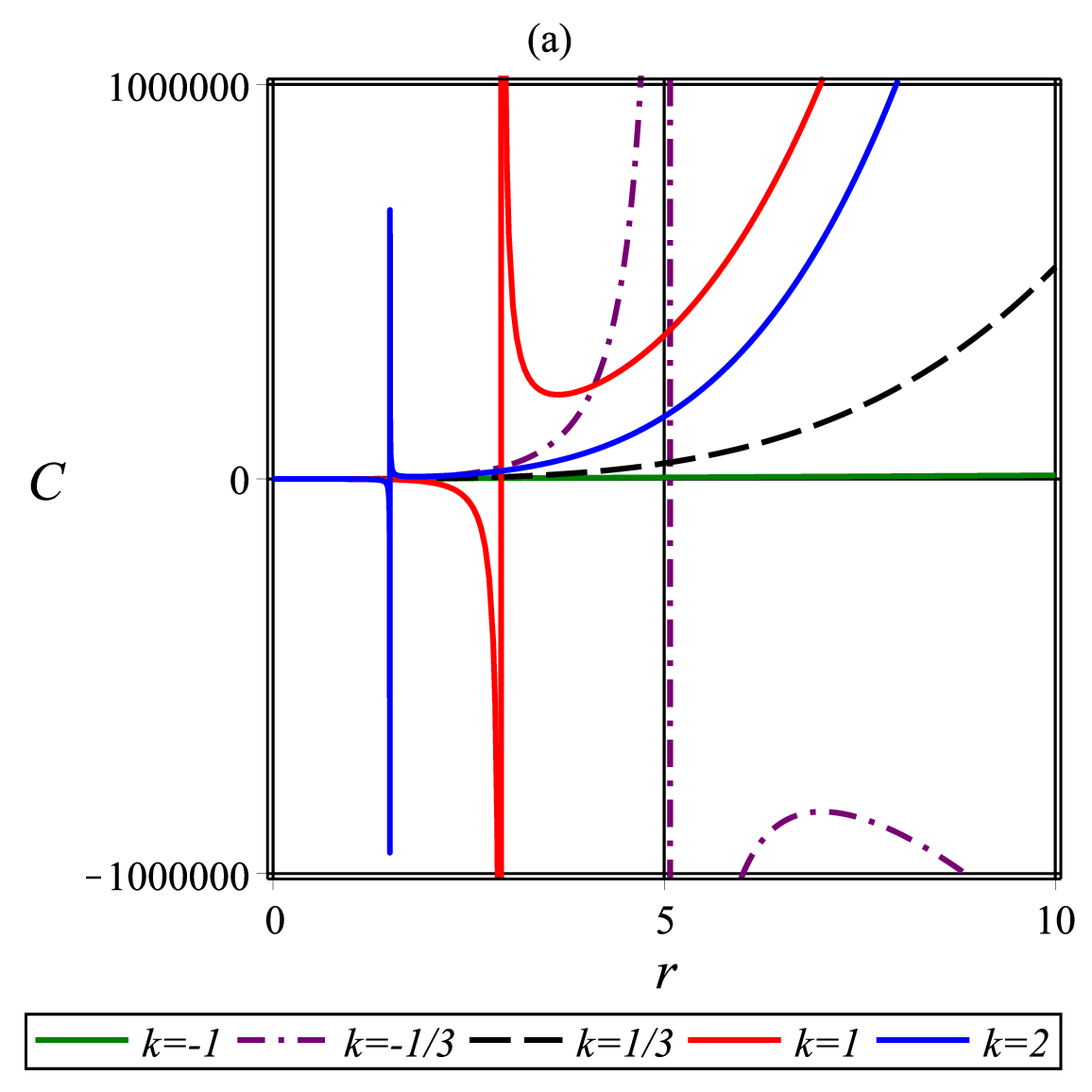}
\includegraphics[scale=0.35]{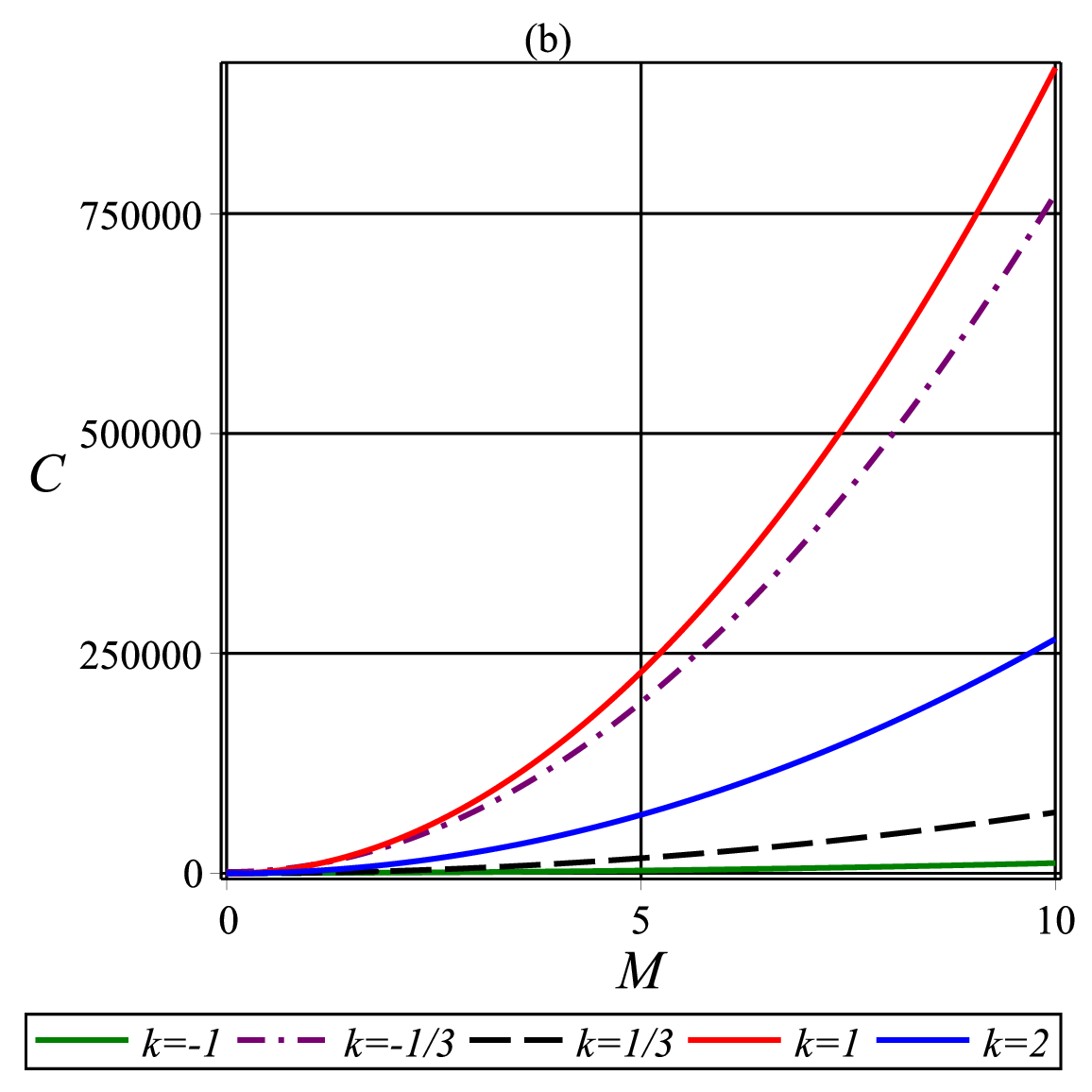}
\caption{\label{fig12} Typical behavior of the
specific heat in terms of (a) radius with $\mathcal{M}=5$
and (b) mass parameter $\mathcal{M}$ with $r_{h}=4$ for
$f_{1}(t)=f_{2}(t)=c=c_{1}=c_{2}=2$.}
\end{figure}

Finally, we can study specific heat in constant volume,
\begin{equation}\label{C}
C=\left(\frac{dU}{dT}\right)_{V},
\end{equation}
which yields to the following expression,
\begin{equation}\label{C2}
C=\pi r_{h}^{2}\frac{kf_{1}r_{h}^{1-2k}-(f_{2}-\frac{1}{2}\mathcal{M}^{2}c_{1}cr_{h}^{2})(k-\frac{1}{2})}{f_{2}(k-\frac{1}{2})-kf_{1}(k+\frac{1}{2})r_{h}^{1-2k}},
\end{equation}
In the plots of the Fig.\ref{fig12} we can see typical behavior of the specific heat in terms of the mass parameter and radius. We can see that specific heat may be positive or negative (for $k$ of order unit) which means some instability with possible phase transition.  One can study such instability in the context of thermal fluctuations \cite{T1,T2,T3,T4,T5,T6} and find that presence of thermal fluctuations may remove mentioned instabilities. Also, Fig.\ref{fig12} (b) shows that specific heat is increasing function of mass parameter.

\section{Conclusions and Discussion}
In this paper, we have analyzed the gravitational collapse in
massive gravity. It was observed that the dynamics of this system
are changed due to the addition of a mass to this system. In this
paper, we first obtained  equation of motion for a time dependent
solution  in  massive gravity. Then barotropic equation of state
was used to further analyze such solutions. Finally, we applied
this solution to a dynamics of gravitational collapse. It was
observed that the gravitational collapse depends on the value of
the mass used to deform this theory.

Contours for $X_{0}$ are obtained against the barometric parameter
$k$ for different values of other parameters like $\mathcal{M}$,
$\alpha$, $\beta$, etc. Various regimes of the fluid content of
the universe has been plotted such as radiation ($k>0$),
pressure-less dust ($k=0$), dark energy ($k<0$), phantom ($k<-1$).
From the figures we see that there is no trajectories in the
negative region. This rules out the existence of black holes as
the end state of collapse in the context of massive gravity with the considered
parameters.

From the first Figs.\ref{fig2ab}, \ref{fig3ab}, \ref{fig4ab},
\ref{fig5ab}, \ref{fig6a}, we see that the contours and surfaces
of $X_{0}$ obtained against the parameters  $\mathcal{M}$,
$\alpha$, $\beta$ and $k$ lie in the positive region. This
confirms the existence of positive roots of equation (\ref{28})
and indicates the end state of the collapse can be a  naked
singularity in the context of massive gravity with the considered
parameters.

In Fig.\ref{fig2ab}(a), we  see that an increase in the value of
massive gravity parameter $\mathcal{M}$ decreases the tendency of
formation of NS. Hence, we observe that the dynamics of the system
gets deformed by the addition of graviton mass to this system. In
Fig.\ref{fig2ab}(b), the $k-X_{0}$ trajectories for different
values of $\alpha$ are obtained and represents that greater values
of $\alpha$ increase the tendency to form NS. In contrast, in
Figs.\ref{fig3ab}(a) and \ref{fig3ab}(b), we observe that greater
values of $c$ and $c_{2}$ decrease the tendency to form the NSs.
In Fig.\ref{fig4ab}(a), the variation of $k-\mathcal{M}$ surfaces
are obtained against $X_{0}$. We see that in the dark energy
regime $(k<-1/3)$ the surface pushes towards the positive
direction of $X_{0}$, accompanied with a decrease in
$\mathcal{M}$, thus showing an increased tendency to form NS. In
Fig.\ref{fig4ab}(b), $k-\alpha$ surfaces are obtained against
$X_{0}$. Here also in the dark energy regime, there is an
increased tendency of NS accompanied by an increase in $\alpha$.
In Fig.\ref{fig5ab}(a), $k-c$ surface is obtained against $X_{0}$.
Here accompanied by a decrease in the value of $c$, we witness an
increased tendency of NS in the dark energy regime. In
Fig.\ref{fig5ab}(b), $k-c_{2}$ surface is obtained against
$X_{0}$. The results obtained are same as that of
Fig.\ref{fig5ab}(a). In Fig.\ref{fig6a}, we see that the $k-\beta$
surface is parallel to the $\beta$ axis around $k=0$ and this
shows that the system is not deformed by $\beta$ and hence the
collapse dynamics does not depend on it in the early universe
regime. But what matters is that the surface remains totally in
the $X_{0}>0$ region. Eventually the surface pushes up towards the
positive axis for $k\leq -0.5$, i.e. in the dark energy regime.
This shows an increased tendency to form naked singularities. In
the figs. 4-6, we see that the surfaces totally lie in the
$X_{0}>0$ region thus showing the possibility of formation of
naked singularities. We have also studied the strength of the
singularity formed and showed that under certain conditions, and concluded that 
strong singularities are possible in massive gravity. This is illustrated in Fig.\ref{fig7}.
It would be interesting to analyze the strength of such  naked singularities, 
for various models, and compare the results of massive gravity with general relativity. 

From the above discussion, we conclude that this work provides
counterexamples of the cosmic censorship conjuncture, which states
that every singularity must be covered by an event horizon, in the
context of massive gravity.  We would like to point out that there
are two forms of the cosmic censorship conjuncture
\cite{conj12}-\cite{conj14}. According to the  strong cosmic
censorship conjuncture, no locally naked singularities can occur.
However, according to the weak cosmic censorship conjuncture,
singularities can be locally naked, but they cannot be  globally
naked. It is also possible to analyze the strength of singularities, and this can be done 
using the Tipler's formalism   \cite{Tipler}-\cite{Maharaj}. 
We have applied this formalism to the massive gravity, and demonstrated 
that it possible to have a strong naked singularity 
in massive gravity.

Finally, we study thermodynamics of the model and calculate some
thermodynamical quantities to investigate effect of mass
parameter. For example, in Figs.\ref{fig9}(b) and \ref{fig12}(b),
we find that the thermalization temperature and specific heat
respectively are  increasing function of $\mathcal{M}$. We also
found some instabilities, corresponding to special values of $k$,
and we found stable/unstable black hole phase transition. For the
future work we would like to focus on the instabilities and
consider effect of thermal fluctuations to see that what can
happen with the instable regions.

\section*{Acknowledgments}

P. Rudra acknowledges University Grants Commission (UGC),
Government of India for providing research project grant (No.
F.PSW-061/15-16 (ERO)). P. Rudra also acknowledges Inter
University Centre for Astronomy and Astrophysics (IUCAA), Pune,
India, for awarding Visiting Associateship. F. Darabi acknowledges
financial support of Azarbaijan Shahid Madani University (No.
S/5749-ASMU) for the Sabbatical Leave, and thanks the hospitality
of ICTP (Trieste) for providing support during the Sabbatical
Leave. Y. Heydarzade acknowledges the support of Azarbaijan Shahid Madani University under the approvement (No. 214/D/24019-ASMU).



\begin{thebibliography}{99}


\bibitem{super}A.G. Riess et al., Astron. J. 116, 1009 (1998)
\bibitem{super1}S. Perlmutter et al., Nature 391 , 51 (1998)
\bibitem{super2}A. G. Riess et al., Astron. J. 118, 2668 (1999)
\bibitem{super0}S. Perlmutter et al., Astrophys. J. 517, 565 (1999)
\bibitem{super4}A. G. Riess et al., Astrophys. J. 560, 49 (2001)
\bibitem{super5} J. L. Tonry et al., Astrophys. J. 594, 1 (2003)
\bibitem{DE}P. J. E. Peebles, and B. Ratra, Rev. Mod. Phys.75, 559 (2003)
\bibitem{DE1}E. J. Copeland, M. Sami, and S. Tsujikawa, Int. J. Mod. Phys.D15, 1753 (2006)
\bibitem{DE2}J. A. Frieman, M. S. Turner, and D. Huterer,  Annual Review of Astronomy and Astrophysics, 46, 385 (2008)
\bibitem{M1}
M. Khurshudyan, B. Pourhassan, A. Pasqua, Can. J. Phys. 93 (2015) 449
\bibitem{2a}
H. van Dam and M. J. G. Veltman, Nucl. Phys. B 22, 397 (1970)

\bibitem{a2}
Y. Iwasaki, Phys. Rev. D 2, 2255 (1970)

\bibitem{M2}
S. Upadhyay, B. Pourhassan, H. Farahani, Phys. Rev. D 95, 106014 (2017)
\bibitem{1}  W. Pauli  and M.  Fierz Helv. Phys. Acta 12, 297 (1939)
\bibitem{2}
M. Fierz Helv. Phys. Acta 12, 3 (1939)



\bibitem{4} A. I. Vainshtein, Phys. Lett. B
39, 393
(1972)
\bibitem{5} E. Babichev and C. De ayet, Class. Quant. Grav.
30, 184001
(2013)
\bibitem{6} D. G. Boulware, S. Deser, Phys. Rev. D
6, 3368
(1972)
\bibitem{7}de  Rham  C,  Gabadadze  G  and  Tolley  A  J  2011
Phys. Rev. Lett.
106,
231101 (2011)
\bibitem{8} de Rham C and Gabadadze G
Phys. Rev.
D
82,
04402 (2010)
\bibitem{9}  de Rham C, Gabadadze G and Tolley A J
Phys. Lett.
B
711,
190 (2012)
\bibitem{10}  S. F. Hassan, R. A. Rosen and A. Schmidt-May, JHEP
1202, 026
(2012)
\bibitem{11}S. F. Hassan, A. Schmidt-May and M. von Strauss, Phys. Lett. B
715, 335
(2012)
\bibitem{12}S. F.  Hassan  and R. A. Rosen
Phys. Rev. Lett.
108,
041101 (2012)
\bibitem{13} S. F.  Hassan S F and R. A. Rosen  JHEP
1204,
123 (2012)
\bibitem{14}  K. Hinterbichler, Rev. Mod. Phys.
84, 671
(2012)
\bibitem{lore}
  A.~H.~Chamseddine and V.~Mukhanov,  JHEP { 1208}, 036 (2012)
  \bibitem{l}  I.~Arraut,
   arXiv:1505.06215 [gr-qc].
  \bibitem{l1}S.~Dengiz,  arXiv:1409.5371 [hep-th]
\bibitem{lore2} G.~Goon, K.~Hinterbichler, A.~Joyce and M.~Trodden,
   JHEP {  1507}, 101 (2015)


   \bibitem{gaus} S. H. Hendi, S. Panahiyan, B. E. Panah,   JHEP 01, 129 (2016)
   \bibitem{gauss} Se. H. Hendi, G. Q. Li, J. X. Mo, S. Panahiyan and B. E. Panah,
    Eur. Phys. J. C 76, 571 (2016)
    \bibitem{phase} S. H. Hendi, S. Panahiyan, B. E. Panah and M. Momennia,
    Ann. der Phys.  528, 819 (2016)

\bibitem{cosm12}  M. Wyman, W. Hu and P. Gratia,   Phys. Rev. D 87, 084046 (2013)
\bibitem{cosm14} M. S. Volkov,  Phys.Rev. D 90, 024028 (2014)

\bibitem{ads}
A. Sinha,   JHEP 1006, 061 (2010)
\bibitem{M3}
J. Sadeghi and B. Pourhassan, JHEP12 (2008) 026
\bibitem{M4}
B. Pourhassan  and J. Sadeghi, Can J Phys 91 (2013) 995
\bibitem{M5}
J. Sadeghi, B. Pourhassan and S. Heshmatian, Advances in High Energy Physics 2013 (2013) 759804

\bibitem{cft}
V. Niarchos,  Fortsch. Phys. 57, 646 (2009)

\bibitem{ee} X. X. Zeng, H. Zhang and  L. F. Li, Phys. Lett. B 756,  170 (2016)
\bibitem{cc} W. J.  Pan and Y. C. Huang,   arXiv:1612.03627
\bibitem{conduct}  L. Alberte and A. Khmelnitsky, Phys. Rev. D 91, 046006 (2015)
\bibitem{time} V. Ziogas, JHEP 09, 114 (2015)
\bibitem{time1}  V. Keranen and P.  Kleinert, Phys. Rev. D 94, 026010 (2016)
\bibitem{sphe} P. C. Vaidya, Curr. Sci. 12, 183 (1943)
\bibitem{sphe1} P. C. Vaidya, Nature 171, 260 (1953)
\bibitem{td12} P. Li, X. Z. Li and X. H. Zhai, Phys. Rev. D 94, 124022 (2016)
\bibitem{td21}
Y. P.  Hu, X. X.  Zeng and  H. Q. Zhang, Phys. Lett. B 765,  120 (2017)
\bibitem{M6}
K. B. Fadafan, B. Pourhassan and J. Sadeghi, Eur. Phys. J. C 71, 1785 (2011)
\bibitem{Elena}
E. Caceres, A. Kundu and D. L. Yang, JHEP 1403, 073 (2014)
\bibitem{gc} M. Sharif and A.  Siddiqa,  Gen. Rel. Grav. 43, 73 (2011)
\bibitem{gc12}  M. Sharif and A. Siddiqa,  Mod. Phys. Lett. A 25, 2831 (2010)
\bibitem{r1} P. Rudra, R. Biswas, U. Debnath,  Astrophys. Space Sci. 335, 505 (2011)
\bibitem{r2} U. Debnath, P. Rudra, R. Biswas,  Astrophys. Space Sci. 339, 135 (2012)
\bibitem{r3} P. Rudra, R. Biswas, U. Debnath,  Astrophys. Space Sci. 354, 597 (2014)
\bibitem{r4} P. Rudra, U. Debnath, Can. J. Phys. 92(11), 1474 (2014)
\bibitem{r5} P. Rudra, M. Faizal, A. F. Ali,  Nucl. Phys. B. 909, 725 (2016)
\bibitem{r6} Y. Heydarzade, P. Rudra, F. Darabi, A. F. Ali, M.
Faizal,  Phys. Lett. B. 774, 46 (2017)

\bibitem{refe1} I.  H.  Dwivedi and P.  S.  Joshi, Class. Quant. Grav. 6, 1599 (1989)
\bibitem{referen}  I.  H.  Dwivedi and P.  S.  Joshi, Class. Quant. Grav. 8, 1339 (1991)
\bibitem{referen1} I.  H.  Dwivedi and P.  S.  Joshi, J. Math. Phys. 32, 2167 (1991)
\bibitem{referen2} I.  H.  Dwivedi and P.  S.  Joshi, Phys. Rev. D 45, 2147 (1992)
\bibitem{reference4} K. Lake, Phys. Rev. D 43, 1416 (1991)
\bibitem{reference4} S.  M.  Wagh and S.  D.  Maharaj, Gen. Rel. Grav. 31, 975 (1999)
\bibitem{reference4}   S.  G.  Ghosh and A.  Beesham, Phys. Rev. D 61, 067502 (2000)
\bibitem{reference4}   S.  G.  Ghosh and N.  Dadhich,, Phys. Rev. D 64, 047501 (2001)
\bibitem{refe2}A.  Beesham and S.  G.  Ghosh,  Int. J. Mod. Phys. D12, 801 (2003)




\bibitem{reference} R. G. Cai, Y. P. Hu, Q. Y. Pan and Y. L. Zhang, Phys. Rev. D 91 024032 (2015)
\bibitem{Eardley1} D. M. Eardley,  and  L. Smar ,  Phys. Rev. D 19,  2239 (1979)
\bibitem{Joshi1} P. S. Joshi, I. H. Dwivedi,  Commun. Math. Phys.  146,  333 (1992)
\bibitem{Joshi2} P. S. Joshi, T. P. Singh,  Phys. Rev. D 51,  6778 (1995)
\bibitem{Singh1} T. P. Singh, P. S. Joshi,  Class. Quant. Grav. 13, 559 (1996)

\bibitem{Tipler} F. J. Tipler, Phys. Lett. A. 64, 8 (1977)
\bibitem{conj15}V. D. Vertogradov, J  Phys. Conf. Series,  769, 012013 (2016)
\bibitem{conj16} V.D. Vertogradov, Grav. Cosmol. 22, 220 (2016)
 \bibitem{Maharaj} M. D. Mkenyeleye, R. Goswami, and S. D. Maharaj, Phys. Rev. D 90,  064034 (2014)



\bibitem{T1}
B. Pourhassan and M. Faizal,  Nucl. Phys. B 913, 834 (2016)
\bibitem{T2}
J. Sadeghi, B. Pourhassan and M. Rostami, Phys. Rev. D 94, 064006 (2016)
\bibitem{T3}
B. Pourhassan and M. Faizal, Phys. Lett. B 755, 444 (2016)
\bibitem{T4}
B. Pourhassan, M. Faizal and U. Debnath, Eur. Phys. J. C 76, 145 (2016)
\bibitem{T5}
M. Faizal and B. Pourhassan, Phys. Lett. B 751, 487 (2015)
\bibitem{T6}
B. Pourhassan and M. Faizal, Europhys. Lett. 111, 40006 (2015)

\bibitem{conj12} T. P. Singh,   gr-qc/9606016
\bibitem{conj14} S. S. Deshingkar, S. Jhingan and  P. S. Joshi, Gen. Rel. Grav. 30, 1477 (1998)




\end{thebibliography}
\end{document}